\documentclass[journal,10pt]{IEEEtran}
\usepackage{graphicx}
\usepackage{verbatim}
\usepackage{upgreek}
\usepackage{amssymb,amsmath}
\usepackage{color}
\usepackage{epstopdf}
\usepackage{bm}
\usepackage{cite}
\usepackage{array,color}
\usepackage{algorithm}
\usepackage{algorithmicx}
\usepackage{algpseudocode}
\usepackage{amsmath}
\usepackage{amsfonts}
\usepackage{amsthm}
\newtheorem{theorem}{Theorem}
\usepackage{graphics}
\usepackage{epsfig}
\usepackage{soul}
\usepackage{booktabs}
\usepackage{multirow}
\usepackage{mathrsfs}
\soulregister\cite7
\soulregister\ref7
\usepackage{subfigure}  
\usepackage{tabularx}
\usepackage{makecell}
\usepackage{graphicx}
\usepackage[font=small]{caption}

\usepackage{etoolbox}
\newcommand{\tabincell}[2]{\begin{tabular}{@{}#1@{}}#2\end{tabular}} 
\makeatletter
\patchcmd{\@makecaption}
{\scshape}
{}
{}
{}
\makeatletter
\patchcmd{\@makecaption}
{\\}
{.\ }
{}
{}
\makeatother

\ifCLASSINFOpdf
\else
\fi

\begin{document}

\title{Wireless Powered MEC Systems via Discrete Pinching Antennas: TDMA versus NOMA}

\author{
	\IEEEauthorblockN{Peng Liu,~Zesong Fei,~\IEEEmembership{Senior~Member,~IEEE},~Meng Hua,~\IEEEmembership{Member,~IEEE},\\~Guangji Chen,~Xinyi Wang,~\IEEEmembership{Member,~IEEE}},~Ruiqi Liu,~\IEEEmembership{Senior~Member,~IEEE}
	
	\thanks{Peng Liu, Zesong Fei, and Xinyi Wang are with the School of Information and Electronics, Beijing Institute of Technology, Beijing 100081, China (e-mail: bit\_peng\_liu@163.com,   feizesong@bit.edu.cn, bit\_wangxy@163.com).}
    \thanks{Meng Hua is with the Department of Electrical and Electronic Engineering, Imperial College London, London SW7 2AZ, UK. (e-mail: m.hua@imperial.ac.uk).}
	\thanks{Guangji Chen is with the School of Electronic and Optical Engineering, Nanjing University of Science and Technology, Nanjing 210094, China, and also with the National Mobile CommunicationsResearch Laboratory, Southeast University, Nanjing, China (e-mail: guangjichen@njust.edu.cn).}
\thanks{R. Liu is with the Wireless and Computing Research Institute, ZTE Corporation, Beijing 100029, China  (e-mails: richie.leo@zte.com.cn).}

}

\maketitle

\begin{abstract}

Pinching antennas (PAs), a new type of reconfigurable and flexible antenna structures, have recently attracted significant research interest due to their ability to create line-of-sight links and mitigate large-scale path loss. Owing to their potential benefits, integrating PAs into wireless powered mobile edge computing (MEC) systems is regarded as \textcolor{black}{a viable solution to enhance both energy transfer and task offloading efficiency.} Unlike prior studies that assume ideal continuous PA placement along waveguides, this paper investigates a practical discrete PA-assisted wireless power MEC framework, where devices first harvest energy from PA-emitted radio-frequency signals and then adopt a partial offloading mode, allocating part of the harvested energy to local computing and the remainder to uplink offloading. The uplink phase considers both the time-division multiple access (TDMA) and non-orthogonal multiple access (NOMA), each examined under three levels of PA activation flexibility. For each configuration, we formulate a joint optimization problem to maximize the total computational bits and conduct a theoretical performance comparison between the TDMA and NOMA schemes. To address the resulting mixed-integer nonlinear problems, we develop a two-layer algorithm that combines closed-form solutions based on Karush–Kuhn–Tucker (KKT) conditions with a cross-entropy-based learning method. \textcolor{black}{Numerical results validate the superiority of the proposed design in terms of the harvested energy and computation performance, revealing that TDMA and NOMA achieve comparable performance under coarser PA activation levels, whereas finer activation granularity enables TDMA to achieve superior computation performance over NOMA.}

\vspace{1ex}
\end{abstract}

\IEEEpeerreviewmaketitle
\section{Introduction}
In the era of sixth-generation (6G) wireless networks, emerging applications such as smart homes, intelligent transportation, and augmented reality are expected to provide immersive, real-time, and ubiquitous services \cite{iot1}. These applications heavily rely on the acquisition of massive sensing data and require significant computational resources to support latency-sensitive and computation-intensive tasks \cite{iot2}. To meet these demands, mobile edge computing (MEC) has emerged as a promising paradigm that brings powerful computing and storage capabilities closer to end devices \cite{MEC1}. By offloading computational tasks from resource-constrained devices to nearby edge servers, MEC can effectively reduce communication latency, enhance energy efficiency, and improve the overall quality of service. However, the benefits of MEC are critically constrained by the limited energy supply of devices, which often operate with finite battery capacity. To overcome this limitation, wireless power transfer (WPT) has been introduced as a key enabling technology \cite{WPT1}, allowing devices to harvest energy from radio-frequency (RF) signals transmitted by the base station (BS). By integrating WPT with MEC, namely WPT-MEC, devices can achieve sustained operation and seamless task offloading, thereby unlocking the full potential of computation-intensive applications in future 6G networks.

The integration of WPT and MEC has garnered considerable research attention.  In \cite{MEC-WPT1}, the authors investigated a time division multiple access (TDMA)-based WPT-MEC system, where transmission beamforming and timeslot allocation were jointly optimized to minimize the BS’s total energy consumption under latency constraints. Building on this work, the authors of \cite{MEC-WPT2} considered a dynamic scenario in which computational tasks arrive over time. By jointly optimizing the BS’s energy allocation across slots and the scheduling of user computations, transmission energy was minimized over a finite horizon.  Furthermore,  the authors of  \cite{MEC-WPT3} examined a non-orthogonal multiple access (NOMA)-based WPT-MEC system and addressed the problem of maximizing the computation energy efficiency. By leveraging fractional programming,  the device transmit powers and MEC server computational frequencies were jointly optimized, thereby achieving significant improvements in overall energy efficiency. Although these studies have demonstrated the potential of WPT-MEC systems, their performance remains highly sensitive to wireless channel conditions. In particular, once the channel quality deteriorates, such as in the presence of obstacles or severe path loss, both the WPT and computation offloading efficiency can be significantly degraded.

Flexible antenna technologies, such as reconfigurable intelligent surfaces (RIS) \cite{RIS1,RIS2}, movable antennas \cite{Mov1,Mov3}, and fluid antennas \cite{fas1,fas2}, have recently emerged as promising paradigms for proactively adjusting wireless channel conditions. Specifically, RIS employs a large array of passive reflecting elements with tunable phase shifts to constructively align signals at the receiver, thereby improving both the energy transfer efficiency and computation offloading performance \cite{RIS3,RIS4}. Movable and fluid antennas leverage spatial repositioning within a confined region to alter instantaneous channel state information (CSI) and mitigate fading effects \cite{mov4}. However, these technologies face inherent limitations. For example, RIS often suffers from double-fading attenuation due to the cascaded BS–RIS–device link, while movable and fluid antennas are typically restricted to displacements of only a few wavelengths, limiting their capability to combat large-scale path loss. These challenges motivate the exploration of alternative architectures, such as pinching antennas (PAs). First introduced by NTT DOCOMO in 2022 \cite{suzuki2022pinching}, PAs represent a novel form of reconfigurable antenna based on leaky-wave radiation through elongated, dielectric-filled waveguides. Discrete “pinched” dielectric inclusions are embedded along the waveguide to locally control signal emission and reception. By selectively activating these pinching points, PAs dynamically steer radiation to establish line-of-sight (LoS) links with nearby devices \cite{PASS1}. A key advantage of PAs lies in their physical flexibility: the waveguide structure can be arbitrarily extended to bring the radiating aperture closer to target devices, thereby significantly reducing propagation loss. Moreover, their hardware simplicity—requiring only the insertion or removal of dielectric particles—makes them a cost-effective and scalable alternative to conventional reconfigurable antennas \cite{PASS4}. Given these properties, PAs are particularly well-suited for WPT-MEC systems, where establishing robust and reconfigurable LoS links is essential for the efficient WPT and task offloading.

Research on PAs still remains in its infancy, with existing studies primarily focusing on their potential in wireless communication systems. For PA-assisted downlink communications, the authors of \cite{PASS0,PassHua,PASS0-1} investigated the rate optimization enabled by PAs, showing that PA technology can achieve higher communication rates compared with conventional antenna setups. To address the issue of CSI acquisition, \cite{PASSce} proposed an efficient CSI estimation scheme that activates only a small number of PAs while achieving accurate multipath parameter estimation. In \cite{PASS5}, the authors derived a closed-form upper bound for the array gain of PAs with fixed antenna spacing and further showed that optimized spacing can yield significant performance gains. The study of \cite{PASS6} identified new challenges in resource allocation for PA-enabled systems, focusing on the joint optimization of PA positioning, power control, timeslot scheduling, and subcarrier allocation, with tailored algorithms proposed to address these issues. In \cite{PASS7}, the fundamental capacity limits of PA-assisted multiple-access channels were examined, and a lower bound on the sum-rate was derived under NOMA transmission. Furthermore, the authors of \cite{PASS8} considered the PA placement optimization in both single- and multi-waveguide scenarios to enhance secrecy performance. For PA-assisted uplink communications, the authors of \cite{UPASS1} investigated the sum-rate maximization problem in a PA-assisted multiuser uplink system and proposed a gradient-descent-based algorithm to jointly optimize PA positions and user transmit powers, thereby enhancing uplink throughput. Furthermore, the authors of \cite{UPASS2} addressed the fairness issue by jointly optimizing PA positions and bandwidth allocation to maximize the minimum user rate.  In \cite{liu2025vtc}, the authors studied a PA-assisted WPT-MEC system under the NOMA framework and proposed a particle swarm optimization algorithm to optimize PA placement for enhancing the system’s computational capacity. It is worth noting that most existing studies assume PA positions can be modeled as continuous variables \cite{PASS5,PASS6,PASS7,PASS8, UPASS1,UPASS2,liu2025vtc}, allowing arbitrary relocation or activation, which is often infeasible in practical deployments due to physical and hardware constraints. A more realistic design is to pre-deploy PAs at discrete candidate positions along the waveguide and selectively activate them to improve system performance. In this context, \cite{dpa1} investigated the antenna activation problem in PA-assisted downlink communications under NOMA transmission and proposed a matching-theory-based method to maximize the overall transmission rate. Extending this work, \cite{dpa2} considered multi-waveguide systems and developed a game-theoretic framework, combined with successive convex approximation, to jointly optimize activation strategies and power allocation. Nevertheless, these studies have mainly focused on conventional downlink communication systems. The investigation of discrete PA activation in energy-constrained and computation-intensive scenarios such as WPT-MEC systems remains largely unexplored. 

A key challenge in discrete PA-assisted WPT-MEC systems lies in the intrinsic coupling between downlink and uplink: the energy delivered in the downlink directly determines the devices’ available computing and transmission power, thereby complicating system design and optimization. Furthermore, most existing studies primarily focused on NOMA-based frameworks with continuous PA placement \cite{PASS0,PASS0-1,PASS7,liu2025vtc}, leaving open the fundamental question of whether NOMA or TDMA is more suited for discrete PA-enabled WPT-MEC scenarios. Another open issue concerns the strategy of dynamic PA activation. In particular, it remains unclear whether identical PA activation strategies should be applied to both the downlink and uplink phases, and whether activating all PAs is always optimal or selectively activating a subset can yield better system performance. 

Motivated by these considerations, this paper investigates a discrete PA-assisted WPT-MEC system, in which multiple PAs are pre-deployed uniformly along the waveguide and selectively activated to enable downlink energy transfer and uplink signal reception. In the downlink phase, devices harvest wireless energy, which is then utilized for both the uplink task offloading and local computation. We consider two multiple-access schemes, namely TDMA and NOMA, with the objective of maximizing the total computational bits by jointly optimizing PA activation strategies, timeslot allocation, transmit power, and computational frequency. The main contributions of this work are summarized as follows:
\begin{itemize}
	\item{We propose a  discrete PA-assisted WPT-MEC architecture, where devices first harvest energy from RF signals transmitted via discrete PAs and then operate in a partial offloading mode, allocating part of the harvested energy to local computing and the remainder to uplink offloading. Both TDMA and NOMA  schemes are considered for uplink offloading, and for each scheme, we explore three levels of dynamic PA activation: \textcolor{black}{\textbf{(i) Static PA activation:}} applying a common activation strategy for both downlink and uplink phases; \textcolor{black}{\textbf{(ii) Partially dynamic PA activation:}} adopting distinct activation strategies for the two phases; \textcolor{black}{\textbf{(iii) Fully dynamic PA activation:}} further allowing different activation patterns across individual uplink sub-slots. For each PA activation configuration, we formulate a joint optimization problem of PA activation, transmit power, computational  frequency, and timeslot allocation, with the objective of maximizing the total number of computed bits.}
	
	\item{We provide a theoretical comparison of computation performance across different schemes and PA activation configurations. Specifically, we first characterize the performance relationships among the three PA activation configurations under both the TDMA and NOMA. By leveraging optimization theory, we further derive the structural properties of the optimal solutions and prove that TDMA and NOMA achieve identical computation performance in \textbf{static} and \textbf{partially dynamic PA activation}, while in \textbf{fully dynamic PA activation}, the TDMA scheme achieves superior computation performance due to its higher flexibility in PA activation.}
	
	\item{To address the resultant mixed-integer nonlinear programming (MINLP) problem, we take the partially dynamic PA activation in TDMA as an example and decouple it into an inner optimization problem over timeslot allocation, transmit power, and computational frequency, and an outer optimization problem over PA activation. An efficient two-layer algorithm is then developed: closed-form solutions are derived for the inner problem using Karush–Kuhn–Tucker (KKT) conditions, while the outer problem is tackled using a cross-entropy-based method to iteratively learn the optimal PA activation distribution. This algorithmic framework is subsequently shown to be extendable to the remaining PA activation configurations under both TDMA and NOMA schemes.}
	\item{Numerical results validate the performance gains of the proposed discrete PA-assisted WPT-MEC system. Compared with conventional antenna-based designs, our approach achieves higher total computational bits and greater WPT efficiency. Furthermore, due to the enhanced phase alignment, selective PA activation outperforms full activation. Consistent with the theoretical analysis, dynamic PA activation plays a critical role in system performance, and TDMA demonstrates superior results over NOMA under more flexible activation strategies.}

\end{itemize}

The remainder of this paper is organized as follows. Section II introduces the system model of the discrete PA-assisted WPT-MEC framework and formulates the corresponding optimization problems for three PA activation configurations under both TDMA and NOMA schemes. Section III presents a theoretical performance analysis and compares different multiple access schemes and activation strategies. In Section IV, a two-layer optimization algorithm is developed based on KKT conditions and the cross-entropy method. Section V provides numerical results to validate the effectiveness of the proposed design. Finally, Section VI concludes the paper.

\section{System Model and Problem Formulation}

\begin{figure}[!t]
	\centering
	\includegraphics[width=2.9in]{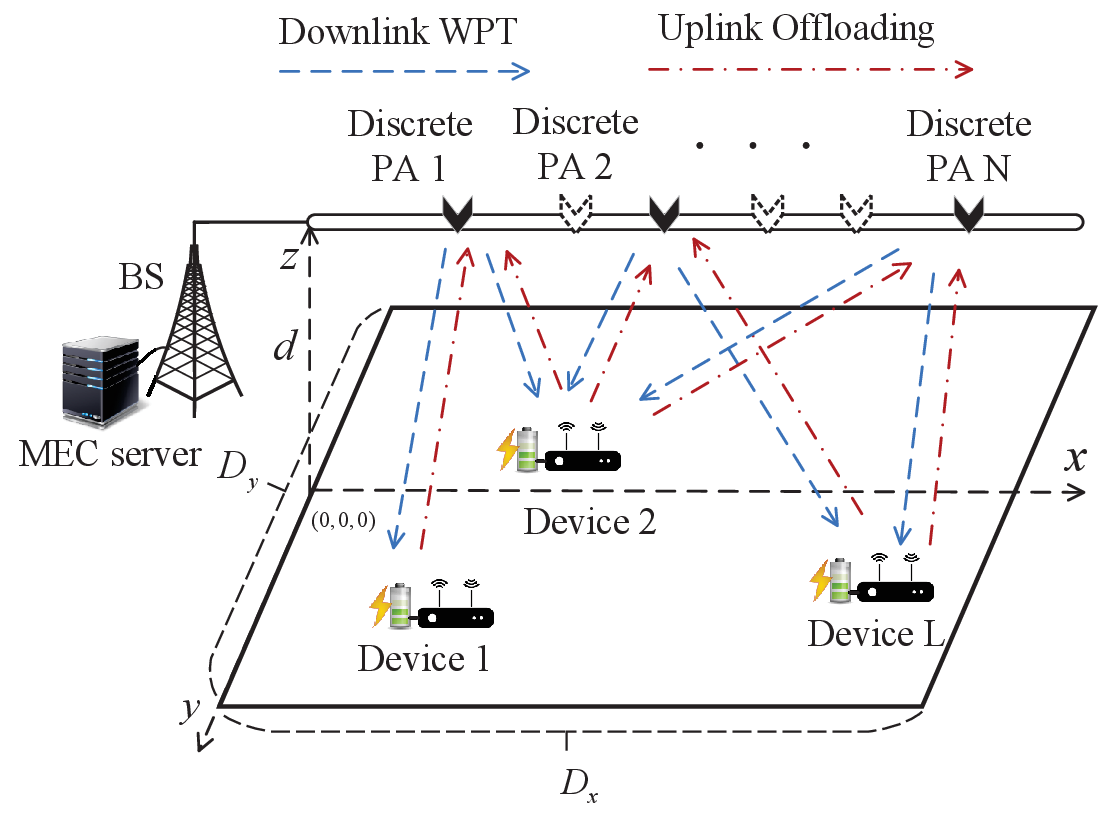}
	\caption{An illustration of discrete PA-enabled  WPT-MEC systems.}
	\label{fig:sys}
\end{figure}

\subsection{System Model}
As illustrated in Fig. 1, we consider a PA-assisted WPT-MEC system consisting of a BS integrated with an MEC server and $L$ single-antenna devices. A waveguide equipped with $N$  PAs is connected to the BS and deployed at a height of $d$~meter (m) to serve the devices. The PAs are uniformly pre-installed along the waveguide and can be dynamically activated\footnote{A practical way to realize discrete pinching antennas is to construct a track parallel to the waveguide and place dielectric pinches at $N$ designated positions. By selectively engaging these pinches with the waveguide, any subset of antennas can be activated.}, where the position of the $n$-th PA is denoted by $\bm{\psi}^{\mathrm{P}}_n=(x_n,0,d)$, and the feed point of the waveguide is located at the reference position $\bm{\psi}^{\mathrm{D}}_0=(0,0,d)$. The devices are assumed to be randomly distributed within a rectangular region of size $D_x \times D_y$, with their coordinates denoted as $\bm{\psi}_l=(x_l,y_l,0), l=1,...,L$. Each device performs its computation-intensive task in a partial offloading mode, where the task is split into two parts: one locally computed and the other offloaded to the MEC server. Moreover, each device is  equipped with a rechargeable battery for energy harvesting. We adopt a typical time-division duplex (TDD)-based “harvest-then-offload” protocol \cite{tdd,RIS3}. Specifically, within a transmission frame $T$, the BS first broadcasts wireless signals via the PAs to charge the devices during the downlink phase duration  $t_0$, after which the harvested energy is used to offload computational tasks to the MEC server in the uplink phase duration  $t_1$.

To enable a flexible antenna activation, we define the PA activation vectors for the downlink and uplink as $\bm{\beta}^{\mathrm{D}}=\{\beta^{\mathrm{D}}_1,\ldots,\beta^{\mathrm{D}}_N\}$ and $\bm{\beta}^{\mathrm{U}}=\{\beta^{\mathrm{U}}_1,\ldots,\beta^{\mathrm{U}}_N\}$, respectively. The activations variables $\beta^{\mathrm{D}}_n,\beta^{\mathrm{U}}_n\in\{0,1\}$ indicate the state of the $n$-th pinching antenna in the downlink and uplink phases, with $\beta^{\mathrm{D}}_n=1$ indicating activation and $\beta^{\mathrm{D}}_n=0$ indicating inactivity. Based on the spherical wave channel model, the downlink WPT channel from the feed point, via the $n$-th PA, to the $l$-th device is expressed as \cite{PASS0,PASS4}
\begin{equation} \label{channel}
h^{\mathrm{D}}_{nl}=\underbrace{\frac{\eta{\beta^{\mathrm{D}}_{n} }}{\left\|{\bm{\psi}_l}-\bm{\psi}^{\mathrm{P}}_n\right\|}}_{\text {free-space path loss }} \cdot \underbrace{e^{-j \frac{2 \pi}{\lambda}\left\|{\bm{\psi}_l}-\bm{\psi}^{\mathrm{P}}_n\right\|}}_{\text {free-space phase shift }} \cdot \underbrace{e^{-j \frac{2 \pi}{\lambda_g}\left\|\bm{\psi}^P_0-\bm{\psi}^{\mathrm{P}}_n\right\|}}_{\text {in-waveguide phase shift }},
\end{equation}
where $\eta=\tfrac{c}{4\pi f_c}$ is the frequency-dependent factor, with $c$ and $f_c$ denoting the speed of light and the carrier frequency, respectively. The terms $\left|{\bm{\psi}_l}-\bm{\psi}^{\mathrm{P}}_n\right|$ and $\left|\bm{\psi}^P_0-\bm{\psi}^{\mathrm{P}}_n\right|$ denote the distance between device $l$ and pinching antenna $n$, and that between pinching antenna $n$ and the feed point, respectively. Here, $\lambda$ is the free-space wavelength, while $\lambda_g=\tfrac{\lambda}{n_e}$ is the guided wavelength, with $n_e$ being the effective refractive index of the dielectric waveguide \cite{PASS0}. Since waveguides incur extremely small transmission loss (e.g., 0.01–0.03 dB/m at 15 GHz for a circular copper waveguide), the attenuation along the waveguide is considered negligible \cite{PASS1,PASS4}. Based on the linear channel model in (\ref{channel}), the downlink PA gain for the $l$-th device is expressed as
\begin{equation}
G^{\mathrm{D}}_l(\bm{\beta}^{\mathrm{D}})={\left| \sum\nolimits_{n=1}^{N}h^{\mathrm{D}}_{nl} \right|}^{2}.
\end{equation}

Moreover, we assume that the BS transmits the energy signal with a constant power $P_b$ during the downlink WPT duration $t_0$, and that each PA follows a practical equal-power radiation model \cite{PASS4}. Based on the energy harvesting model in \cite{MEC-WPT1,MEC-WPT2}, the energy harvested by device $l$ is expressed as\footnote{\textcolor{black}{Although the nonlinear model captures energy harvesting more precisely, under low-energy conditions its performance is nearly indistinguishable from the linear model. For analytical tractability, we adopt the linear model in this work, while noting that the proposed framework is sufficiently general to accommodate nonlinear harvesting as well.}}
\begin{equation}
{{E}_{l}}=\gamma t_0 \frac{{P}_{b}}{||\bm{\beta}^{\mathrm{D}}||_0}G^{\mathrm{D}}_l(\bm{\beta}^{\mathrm{D}}),\forall l,
\end{equation}
where $\gamma\in(0,1]$ is the energy conversion efficiency and $||\bm{\beta}^{\mathrm{D}}||_0$ denotes the number of activated PAs in the downlink WPT phase.

In the uplink offloading phase, the channel from the $l$-th device, through the $n$-th PA, to the feed point is given by
\begin{equation}
h^{\mathrm{U}}_{nl}=\frac{\eta \beta_n^{\mathrm{U}} e^{-j \frac{2 \pi}{\lambda}\left\|{\bm{\psi}_l}-\bm{\psi}^{\mathrm{P}}_n\right\|} \cdot e^{-j \frac{2 \pi}{\lambda_g}\left\|\bm{\psi}^{\mathrm{P}}_0-\bm{\psi}^{\mathrm{U}}_n\right\|}}{\left\|{\bm{\psi}_l}-\bm{\psi}^{\mathrm{U}}_n\right\|}.
\end{equation}

The uplink PA gain corresponding to the $l$-th device is given by
\begin{equation}
G^{\mathrm{U}}_l(\bm{\beta}^{\mathrm{U}})={\left| \sum\nolimits_{n=1}^{N}h^{\mathrm{U}}_{nl} \right|}^{2}.
\end{equation}

\begin{figure}[!t]
	\centering
	\includegraphics[width=2.9in]{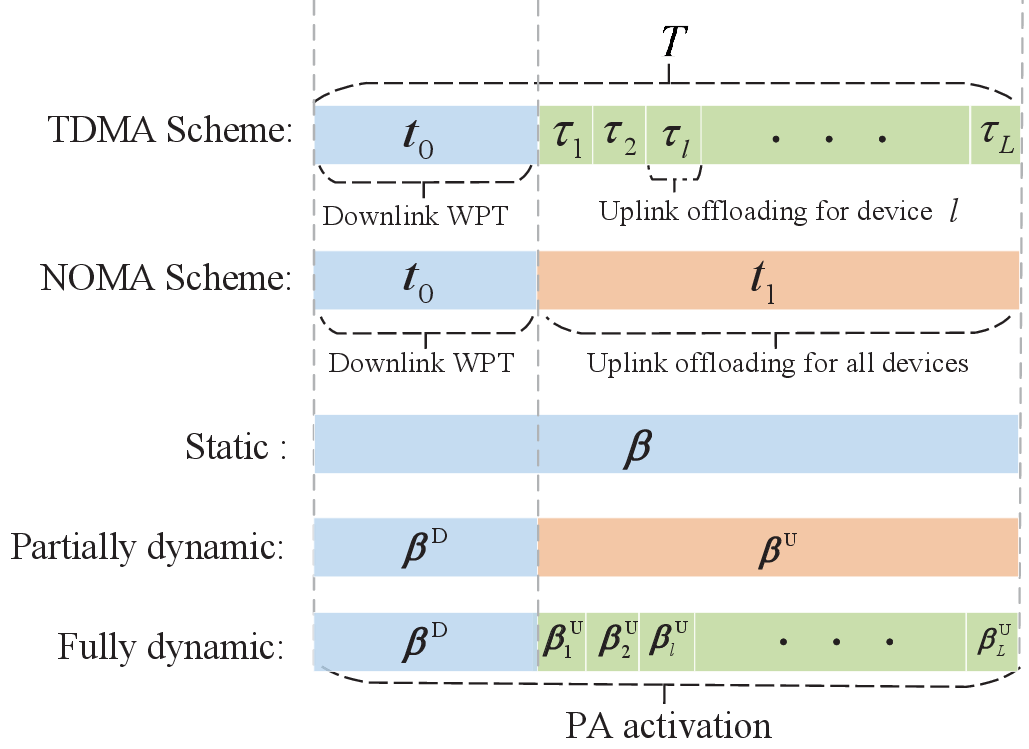}
	\caption{\textcolor{black}{An Illustration of  a transmission frame.}}
	\label{fig:slot}
\end{figure}

As shown in Fig.~\ref{fig:slot}, we assume that devices offload their computational tasks over the same frequency band using either TDMA or NOMA. According to the level of dynamic PA activation, three PA activation configurations are considered:
\begin{itemize}
	\item \textbf{Static PA activation:} A fixed PA activation strategy is applied throughout the entire transmission frame.
	\item \textbf{Partially dynamic PA activation:} Different PA activation strategies are adopted for the downlink and uplink, while all devices share the same PA activation strategy during the uplink.
	\item \textbf{Fully dynamic PA activation:} Flexible PA activation strategies are allowed not only between the downlink and uplink, but also across the uplink offloading phases of different devices.
\end{itemize}

For the TDMA scheme, the uplink offloading duration $t_1$ is further divided into $L$ sub-slots, denoted by $\{\tau_1, \ldots, \tau_L\}$ with $\sum_{l=1}^{L}\tau_l = t_1$, where the $l$-th sub-slot $\tau_l$ is allocated to device $l$ for task offloading. \color{black}We next present the total offloading  bits for each PA activation configuration within a transmission frame $T$.

\textbf{(1) Static PA activation:} Since the same antenna activation strategy $\bm{\beta}$ is applied to both downlink and uplink phases, the sum offloading  bits  of all devices is given by
\begin{equation}
R^{\mathrm{S}}_{\mathrm{TDMA}}=B \sum_{l=1}^L \tau_{l} \log _2\left(1+\frac{p_lG^{\mathrm{U}}_l(\bm{\beta})}{||\bm{\beta}||_0\sigma_B^2}\right),
\end{equation}
where $B$, $p_l$, and $\sigma_B^2$ denote the signal bandwidth, the transmit power of device $l$, and the power of the additive white Gaussian noise (AWGN) at the BS, respectively. 

\textbf{(2) Partially dynamic PA activation:} Different PA activation strategies, $\bm{\beta}^{\mathrm{D}}$ and $\bm{\beta}^{\mathrm{U}}$, are adopted for the downlink and uplink phases, respectively.  The sum offloading  bits  of all devices is then expressed as
\begin{equation}
R^{\mathrm{PD}}_{\mathrm{TDMA}}=B \sum_{l=1}^L \tau_{l} \log _2\left(1+\frac{p_lG^{\mathrm{U}}_l(\bm{\beta}^{\mathrm{U}})}{||\bm{\beta}^{\mathrm{U}}||_0\sigma_B^2}\right).
\end{equation}

\textbf{(3) Fully dynamic PA activation:} Let the uplink PA activation strategy for device $l$ be denoted by $\bm{\beta}_l^{\mathrm{U}}=\{\beta^{\mathrm{U}}_{1,l},\ldots,\beta^{\mathrm{U}}_{N,l}\}$. The sum offloading  bits of all devices can be written as
\begin{equation}
R^{\mathrm{FD}}_{\mathrm{TDMA}}=B \sum_{l=1}^L \tau_{l} \log _2\left(1+\frac{p_lG^{\mathrm{U}}_l(\bm{\beta}_l^{\mathrm{U}})}{||\bm{\beta}_l^{\mathrm{U}}||_0\sigma_B^2}\right).
\end{equation}

\color{black}For the NOMA scheme, all devices simultaneously offload their computational tasks to the BS during the entire uplink offloading duration $t_1$. To mitigate inter-device interference, the BS employs successive interference cancellation (SIC) to decode the data of each device.  \color{black}The  offloading bits for three PA activation configurations under NOMA are presented as follows:

\textbf{(1) Static PA activation:} The sum offloading bits of all devices under static PA activation are given by \cite{Wuwcl}
\begin{equation}
R^{\mathrm{S}}_{\mathrm{NOMA}}=B t_1 \log _2\left(1+\frac{\sum_{l=1}^Lp_lG^{\mathrm{U}}_l(\bm{\beta})}{||\bm{\beta}||_0\sigma_B^2}\right),
\end{equation}

\textbf{(2) Partially dynamic PA activation:} The sum offloading bits under partially dynamic PA activation are expressed as
\begin{equation}
R^{\mathrm{PD}}_{\mathrm{NOMA}}=B t_1 \log _2\left(1+\frac{\sum_{l=1}^Lp_lG^{\mathrm{U}}_l(\bm{\beta}^{\mathrm{U}})}{||\bm{\beta}^{\mathrm{U}}||_0\sigma_B^2}\right),
\end{equation}

\textbf{(3) Fully dynamic PA activation:} To ensure a fair comparison with the TDMA scheme, the uplink offloading duration is also divided into $L$ sub-slots, denoted by $\tau_k,~k=1,\ldots,L$, where each sub-slot adopts a distinct PA activation strategy, denoted by $\bm{\beta}_k^{\mathrm{U}},~k=1,\ldots,L$. The corresponding total number of  offloading bits is expressed as
\begin{equation}
R^{\mathrm{FD}}_{\mathrm{NOMA}}=B \sum_{k=1}^{L} \tau_k \log _2\left(1+\frac{\sum_{l=1}^Lp_lG^{\mathrm{U}}_l(\bm{\beta}_k^{\mathrm{U}})}{||\bm{\beta}_k^{\mathrm{U}}||_0\sigma_B^2}\right).
\end{equation}

\color{black}We assume that the MEC server has sufficient computational capability, such that the computation time at the server can be regarded as negligible  \cite{tdd,RIS3}. Moreover, \textcolor{black}{since the downlink data rate is generally high owing to the relatively large BS transmit power and favorable channel conditions, and the computation results are much smaller than the input data, the downloading time can be considered negligible \cite{MEC-WPT1,MEC-WPT2}.} Let $f_l$ denote the computational frequency (in central processing unit (CPU) cycles/s) of device $l$. The number of locally computed bits and the corresponding energy consumption are given by
\begin{equation}
C^\text{L}_l = \frac{Tf_l}{I_c},\quad~e^{\text{L}}_l = T\kappa f^3_l, ~~~\forall l,
\end{equation}
where $I_c$ (in cycles/bit) represents the computation intensity, i.e., the number of CPU cycles required to process one bit of raw data, and $\kappa$ is the effective hardware-related power coefficient \cite{tdd}.

\subsection{Problem Formulation}
Based on the above system model, the total number of  computational bits within a transmission frame $T$ consists of both offloaded bits and locally computed bits, both of which are determined by the amount of harvested energy. Specifically, higher harvested energy increases the local computational frequency $f_l$ as well as the transmit power $p_l$, thereby increasing the overall computation capability of each device. By designing effective PA activation strategies, the path loss between the transmitting antennas and the devices can be reduced, which improves WPT efficiency and strengthens uplink channel conditions, thus increasing the offloading rate. Moreover, there exists a fundamental trade-off in the allocation of downlink and uplink phase durations: allocating more time to the downlink improves energy harvesting but reduces the uplink duration for task offloading, potentially decreasing the total computational bits. Therefore, in this paper, we aim to maximize the total number of computational bits by jointly optimizing the PA activation strategies $\bm{\beta}^{\mathrm{D}}$ and $\bm{\beta}^{\mathrm{U}}$, the allocation of downlink and uplink phase durations $t_0$ and $t_1$, as well as the transmit power $p_l$ and computational frequency $f_l$ of the devices. In the following, we formulate the optimization problems for the three PA activation configurations under both TDMA and NOMA schemes.
\subsubsection{TDMA Scheme}
For the TDMA scheme, additional uplink sub-slot allocation variables $\{\tau_1, \ldots, \tau_L\}$ must  be considered. For \textbf{static PA activation}, the same antenna activation scheme is applied to both the uplink and downlink phases, and the corresponding optimization problem is formulated as follows
\begin{align} 
\label{prob29}
\left(\text{P{1.1}}\right):  \max_{\substack{\bm{\beta}, t_0, \{p_l\},\\ \{\tau_l\}, \{f_l\}}} & ~R^{\mathrm{S}}_{\mathrm{TDMA}}+\sum_{l=1}^{L}C^\text{L}_l\\
\text{s.t.}~~
&\beta_n \in\{0,1\},~ \forall n, \tag{\ref{prob29}a}\\
& p_l\tau_l+e^{\text{L}}_l\leq E_l, ~~\forall l, \tag{\ref{prob29}b}\\
& t_0+\sum_{l=1}^{L}\tau_l \leq T,  \tag{\ref{prob29}c}\\
& f_l, p_l, t_0, \tau_l \geq 0, ~~\forall l, \tag{\ref{prob29}d}
\end{align}
where constraint (\ref{prob29}\text{a}) denotes that the PA can be activated arbitrarily, (\ref{prob29}\text{b}) ensures that the energy consumption of each device does not exceed its harvested energy\footnote{\textcolor{black}{{Following \cite{MEC-WPT1,MEC-WPT2}, we assume that at the beginning of each frame, each device is equipped with sufficiently large energy storage, ensuring that the stored energy is never depleted during the frame and is fully replenished by its end.}}}, and  (\ref{prob29}\text{c}) restricts the total length of the transmission frame. For \textbf{partially dynamic PA activation}, different antenna activation strategies are applied to the uplink and downlink phases, and the corresponding optimization problem is formulated as
\begin{align} 
\label{prob30}
\left(\text{P{1.2}}\right):  \max_{\substack{\bm{\beta}^{\mathrm{D}},\bm{\beta}^{\mathrm{U}}, t_0,\\  \{p_l\},\{\tau_l\}, \{f_l\}}} & ~R^{\mathrm{PD}}_{\mathrm{TDMA}}+\sum_{l=1}^{L}C^\text{L}_l\\
\text{s.t.}~~
&\beta^{\mathrm{D}}_n,\beta^{\mathrm{U}}_n \in\{0,1\},~ \forall n, \tag{\ref{prob30}a}\\
&  \text{(\ref{prob29}\text{b})}, \text{(\ref{prob29}\text{c})},\text{(\ref{prob29}\text{d})}.\notag
\end{align}

For \textbf{fully dynamic PA activation}, distinct antenna activation strategies are applied not only between the uplink and downlink phases but also across the uplink sub-slots of different devices. The corresponding optimization problem can be formulated as

\begin{align} 
\label{prob31}
\left(\text{P{1.3}}\right):  \max_{\substack{\bm{\beta}^{\mathrm{D}},\{\bm{\beta}_l^{\mathrm{U}}\}, t_0,\\  \{p_l\},\{\tau_l\}, \{f_l\}}} & ~R^{\mathrm{FD}}_{\mathrm{TDMA}}+\sum_{l=1}^{L}C^\text{L}_l\\
\text{s.t.}~~
&\beta^{\mathrm{D}}_{n},\beta^{\mathrm{U}}_{n,l} \in\{0,1\},~ \forall n,l, \tag{\ref{prob31}a}\\
&  \text{(\ref{prob29}\text{b})}, \text{(\ref{prob29}\text{c})},\text{(\ref{prob29}\text{d})}. \notag
\end{align}
\subsubsection{NOMA Scheme}
Analogous to the TDMA scheme, the optimization problems for the three PA activation configurations under the NOMA scheme are formulated as follows:
\begin{align} 
\label{prob32}
\left(\text{P{2.1}}\right):  \max_{\substack{\bm{\beta}, t_0, t_1, \\ \{p_l\}, \{f_l\}}} & ~R^{\mathrm{S}}_{\mathrm{NOMA}}+\sum_{l=1}^{L}C^\text{L}_l\\
\text{s.t.}~~
&\beta_n \in\{0,1\},~ \forall n, \tag{\ref{prob32}a}\\
& p_lt_1+e^{\text{L}}_l\leq E_l, ~~\forall l, \tag{\ref{prob32}b}\\
& t_0+t_1\leq T,  \tag{\ref{prob32}c}\\
& f_l, p_l, t_0, t_1 \geq 0, ~~\forall l, \tag{\ref{prob32}d}
\end{align}
\begin{align} 
\label{prob33}
\left(\text{P{2.2}}\right):  \max_{\substack{\bm{\beta}^{\mathrm{D}},\bm{\beta}^{\mathrm{U}}, t_0,t_1\\  \{p_l\}, \{f_l\}}} & ~R^{\mathrm{PD}}_{\mathrm{NOMA}}+\sum_{l=1}^{L}C^\text{L}_l\\
\text{s.t.}~~
&\beta^{\mathrm{D}}_n,\beta^{\mathrm{U}}_n \in\{0,1\},~ \forall n, \tag{\ref{prob33}a}\\
&  \text{(\ref{prob32}\text{b})}, \text{(\ref{prob32}\text{c})},\text{(\ref{prob32}\text{d})}. \notag
\end{align}
\begin{align} 
\label{prob34}
\left(\text{P{2.3}}\right):  \max_{\substack{\bm{\beta}^{\mathrm{D}},\{\bm{\beta}_l^{\mathrm{U}}\}, t_0,t_1,\\  \{p_l\}, \{f_l\}}} & ~R^{\mathrm{FD}}_{\mathrm{NOMA}}+\sum_{l=1}^{L}C^\text{L}_l\\
\text{s.t.}~~
&\beta^{\mathrm{D}}_{n},\beta^{\mathrm{U}}_{n,l} \in\{0,1\},~ \forall n,l, \tag{\ref{prob34}a}\\
&  \text{(\ref{prob32}\text{b})}, \text{(\ref{prob32}\text{c})},\text{(\ref{prob32}\text{d})}. \notag
\end{align}

\section{TDMA and NOMA: Performance Analysis for Discrete PA-assisted WPT-MEC}
In this section, we theoretically analyze and compare the computation performance of the WPT-MEC system with discrete PAs under three PA activation configurations for both TDMA and NOMA schemes. We first present two theorems that characterize the computation performance relationships among the three PA activation configurations in TDMA and NOMA, respectively, and reveal the impact of dynamic antenna activation strategies on both schemes. We then conduct a comparative analysis between TDMA and NOMA to identify which scheme is more favorable for the WPT-MEC system with discrete PAs. \textcolor{black}{For clarity, we denote the optimal objective value of the optimization problems under TDMA and NOMA as $O^{a}_{\mathrm{TDMA}}$ and $O^{a}_{\mathrm{NOMA}}$, respectively, where $a=\{\mathrm{S},\mathrm{PD},\mathrm{FD}\}$ represents static PA activation, partially dynamic PA activation, and fully dynamic PA activation, respectively.}

For the comparison of computation performance among the three PA activation configurations under TDMA, we have the following theorem.
\begin{theorem}
In the WPT-MEC system with discrete PAs, when devices adopt TDMA for uplink task offloading, it always holds that $O^{\mathrm{S}}_{\mathrm{TDMA}} \leq O^{\mathrm{PD}}_{\mathrm{TDMA}}  \leq O^{\mathrm{FD}}_{\mathrm{TDMA}} $.
\end{theorem}
\begin{IEEEproof}
When the additional constraint $\bm{\beta}^{\mathrm{D}}=\bm{\beta}^{\mathrm{U}}$ is imposed, problem (P1.2) reduces to problem (P1.1). Since the objectives of PA activation differ between downlink (for energy transfer) and uplink (for task offloading), $\bm{\beta}^{\mathrm{D}}$ and $\bm{\beta}^{\mathrm{U}}$ are not necessarily identical. Therefore, under this constraint, we have $O^{\mathrm{S}}_{\mathrm{TDMA}} \leq O^{\mathrm{PD}}_{\mathrm{TDMA}}$. Similarly, imposing the constraint $\bm{\beta}^{\mathrm{U}}_1=\bm{\beta}^{\mathrm{U}}_2=\cdots=\bm{\beta}^{\mathrm{U}}_L$ reduces problem (P1.3) to problem (P1.2). Furthermore, due to the different locations of devices, the optimal uplink PA activation strategies for different devices are not necessarily identical. Therefore, we have $O^{\mathrm{S}}_{\mathrm{TDMA}} \leq O^{\mathrm{PD}}_{\mathrm{TDMA}}  \leq O^{\mathrm{FD}}_{\mathrm{TDMA}} $.
\end{IEEEproof}

For the three PA activation configurations under NOMA, we have the following theorem.
\begin{theorem}
	In the WPT-MEC system with discrete PAs, when devices adopt NOMA for uplink task offloading, it always holds that $O^{\mathrm{S}}_{\mathrm{NOMA}} \leq O^{\mathrm{PD}}_{\mathrm{NOMA}} = O^{\mathrm{FD}}_{\mathrm{NOMA}}$.
\end{theorem}
\begin{IEEEproof}
Following the proof of Theorem 1, when the additional constraints $\bm{\beta}^{\mathrm{D}}=\bm{\beta}^{\mathrm{U}}$ and $\bm{\beta}^{\mathrm{U}}_1=\bm{\beta}^{\mathrm{U}}_2=\cdots=\bm{\beta}^{\mathrm{U}}_L$ are imposed, problems (P2.3) and (P2.2) reduce to problems (P2.2) and (P2.1), respectively. Hence, $O^{\mathrm{S}}_{\mathrm{NOMA}} \leq O^{\mathrm{PD}}_{\mathrm{NOMA}} \leq O^{\mathrm{FD}}_{\mathrm{NOMA}}$. Furthermore, let the optimal uplink PA activation strategy of problem (P2.3) be denoted by $\bm{\beta}^{\mathrm{U}*}_l,l\in\{1,...,L\}$. Defining the uplink PA activation strategy of problem (P2.2) as $\bm{\beta}^{\mathrm{U}*}=\operatorname*{arg\,max}\limits_{\bm{\beta}_k^{\mathrm{U}},k\in\{1,...,L\}}\frac{\sum_{l=1}^Lp_lG^{\mathrm{U}}_l(\bm{\beta}_k^{\mathrm{U}})}{||\bm{\beta}_k^{\mathrm{U}}||_0}$.  We then obtain the following inequality:
\begin{equation}
\begin{aligned}
 O^{\mathrm{FD}}_{\mathrm{NOMA}} \leq & B \sum_{k=1}^{L} \tau_k \log _2\left(1+\frac{\sum_{l=1}^Lp_lG^{\mathrm{U}}_l(\bm{\beta}^{\mathrm{U}*})}{||\bm{\beta}^{\mathrm{U}*}||_0\sigma_B^2}\right) +\sum_{l=1}^{L}C^\text{L}_l\\
= & B t_1 \log _2\left(1+\frac{\sum_{l=1}^Lp_lG^{\mathrm{U}}_l(\bm{\beta}^{\mathrm{U}*})}{||\bm{\beta}^{\mathrm{U}*}||_0\sigma_B^2}\right)+\sum_{l=1}^{L}C^\text{L}_l, \\
=&O^{\mathrm{PD}}_{\mathrm{NOMA}}.
\end{aligned}
\end{equation}

Since  $O^{\mathrm{PD}}_{\mathrm{NOMA}} \leq O^{\mathrm{FD}}_{\mathrm{NOMA}}$ has already been established, it follows that $O^{\mathrm{PD}}_{\mathrm{NOMA}}=O^{\mathrm{FD}}_{\mathrm{NOMA}}$. Combining this with the previous result, we conclude that $O^{\mathrm{S}}_{\mathrm{NOMA}} \leq O^{\mathrm{PD}}_{\mathrm{NOMA}} = O^{\mathrm{FD}}_{\mathrm{NOMA}}$.
\end{IEEEproof}

Based on the above analysis, we observe that the effect of dynamically adjusting PA activation strategies differs between TDMA and NOMA. Compared with a fixed strategy, using different PA activation strategies in the downlink and uplink phases increases the total computational bits in both schemes. When dynamic adjustment is further extended to allow different PA activation strategies across uplink sub-slots, the computation performance of TDMA can be further improved, whereas NOMA remains unaffected. We now compare the computation performance of TDMA and NOMA by establishing the following theorem.
\begin{theorem}
	 \textcolor{black}{Problems (P1.2) and (P2.2) (or Problems (P1.1) and (P2.1))  yield identical optimal objective values, i.e., $O^{\mathrm{PD}*}_{\mathrm{TDMA}}= O^{\mathrm{PD}*}_{\mathrm{NOMA}}$  (or $O^{\mathrm{S}*}_{\mathrm{TDMA}}= O^{\mathrm{S}*}_{\mathrm{NOMA}}$).} Let the optimal solution of problem (P1.2) be denoted by $\{\bm{\beta}^{\mathrm{D}*},\bm{\beta}^{\mathrm{U}*}, t^*_0, \tau^*_l,p^*_l, f^*_l\}^{\mathrm{PD}}_{\mathrm{TDMA}}$. Then, the corresponding optimal solution $\{\bm{\beta}^{\mathrm{D}*},\bm{\beta}^{\mathrm{U}*}, t^*_0, t^*_1,p^*_l, f^*_l\}^{\mathrm{PD}}_{\mathrm{NOMA}}$ of problem (P2.2) can be obtained from the following relation
\begin{equation}
\begin{aligned}
\bm{\beta}^{\mathrm{D}*}=\bm{\beta}^{\mathrm{D}*},~~~\bm{\beta}^{\mathrm{U}*}=\bm{\beta}^{\mathrm{U}*},~~~t^*_0=t^*_0,\\t^*_1=\sum_{l=1}^{L}\tau^*_l,~~~ p^*_l=p^*_l,~~~f^*_l=f^*_l.
\end{aligned}
\end{equation}
\end{theorem}
\begin{IEEEproof}
	We first prove that $O^{\mathrm{PD}*}_{\mathrm{TDMA}} \leq O^{\mathrm{PD}*}_{\mathrm{NOMA}}$. For notational convenience, let $\Gamma_l=\tfrac{G^{\mathrm{U}}_l(\bm{\beta}^{\mathrm{U}})}{|\bm{\beta}^{\mathrm{U}}|0\sigma_B^2}$ and define $e^\text{o}_l \triangleq p_l\tau_l$. Since each device fully utilizes its harvested energy, constraints (\ref{prob29}b) and (\ref{prob32}b) always hold with equality, i.e., $e^\text{o}_l=E_l-e^{\text{L}}_l$. We begin by analyzing the sub-slot allocation property in the TDMA scheme, under which problem (1.2) can be simplified as
\begin{align} 
\label{prob29_2}
  \max_{{ t_0, \{\tau_l\}}} & ~B \sum_{l=1}^L \tau_{l} \log _2\left(1+\frac{e^\text{o}_l\Gamma_l}{\tau_l}\right)\\
\text{s.t.}~
&~ \text{(\ref{prob29}\text{c})},  \notag
\end{align}
which is a convex problem. Its Lagrangian function is given by
\begin{equation}
\begin{aligned}
\mathcal{L}=B  \sum_{l=1}^{L} \tau_{l} \log _2\left(1+\frac{e^\text{o}_l\Gamma_l}{\tau_l}\right) 
 +\lambda\left(T-t_0-\sum_{l=1}^L \tau_{l}\right),
\end{aligned}
\end{equation}
where $\lambda\geq 0$ is the Lagrange multiplier. We next introduce the auxiliary variable $z_l \triangleq \tfrac{e^\text{o}_l\Gamma_l}{\tau_l}$, which represents the received signal-to-noise ratio (SNR) from device $l$ at the BS. Based on the KKT conditions, we obtain
\begin{equation}
\label{u1}
\frac{\partial \mathcal{L}}{\partial \tau_{l}} 
= B \left[ \log_2(1+z_l) - \frac{z_l}{(1+z_l)\ln 2} \right] - \lambda = 0, \forall l.
\end{equation}

Since the second term in (\ref{u1}) is a monotonically increasing function over $[0,\infty)$ and $\lambda$ is the same for all $l$, it follows that $z_1=z_2=...=z_L$, which implies $\tfrac{e^\text{o}_1\Gamma_1}{\tau_1}=\tfrac{e^\text{o}_2\Gamma_2}{\tau_2}=\cdots=\tfrac{e^\text{o}_L\Gamma_L}{\tau_L}$. Consequently, at the optimality of problem (1.2), the SNR of the received signals is identical across all devices in each sub-slot. Therefore, for the optimal solution $\{\bm{\beta}^{\mathrm{D}*},\bm{\beta}^{\mathrm{U}*}, t^*_0, \tau^*_l,p^*_l, f^*_l\}^{\mathrm{PD}}_{\mathrm{TDMA}}$, the objective value of problem (1.2) can be written as

\begin{align}\label{u2}
O^{\mathrm{PD}*}_{\mathrm{TDMA}}&=B \sum_{l=1}^L \tau^*_{l} \log _2\left(1+\frac{e^\text{o*}_l\Gamma_l}{\tau^*_l}\right)+\sum_{l=1}^{L}C^\text{L*}_l \notag\\
&=B t^*_{1} \log _2\left(1+\frac{\sum_{l=1}^Le^\text{o*}_l\Gamma_l}{\sum_{l=1}^L\tau^*_l}\right)+\sum_{l=1}^{L}C^\text{L*}_l\notag\\
&=B t^*_{1} \log _2\left(1+\frac{\sum_{l=1}^Le^\text{o*}_l\Gamma_l}{t^*_1}\right)+\sum_{l=1}^{L}C^\text{L*}_l.
\end{align}

 From (\ref{u2}), we observe that the optimal solution of problem (1.2) is also a feasible solution of problem (2.2), with $t^*_1=\sum_{l=1}^{L}\tau^*_l$. Therefore, it follows that $O^{\mathrm{PD}*}_{\mathrm{TDMA}} \leq O^{\mathrm{PD}*}_{\mathrm{NOMA}}$.
 
 We next prove the inequality $O^{\mathrm{PD}*}_{\mathrm{TDMA}} \geq O^{\mathrm{PD}*}_{\mathrm{NOMA}}$. For the optimal solution $\{\bm{\beta}^{\mathrm{D}*},\bm{\beta}^{\mathrm{U}*}, t^*_0, t^*_1,p^*_l, f^*_l\}^{\mathrm{PD}}_{\mathrm{NOMA}}$ of problem (2.2), we can always construct a sub-slot allocation scheme  $\sum_{l=1}^{L}\tau^*_l=t^*_1$ that satisfies $\tfrac{e^\text{o}_1\Gamma_1}{\tau_1}=\tfrac{e^\text{o}_2\Gamma_2}{\tau_2}=\cdots=\tfrac{e^\text{o}_L\Gamma_L}{\tau_L}$. With this construction, the objective value of problem (2.2) can be rewritten as

 \begin{align}
  \label{u3}
 O^{\mathrm{PD}*}_{\mathrm{NOMA}}&
 =B t_1 \log _2\left(1+\frac{\sum_{l=1}^Lp^*_lG^{\mathrm{U}}_l(\bm{\beta}^{\mathrm{U}*})}{||\bm{\beta}^{\mathrm{U}*}||_0\sigma_B^2}\right)+\sum_{l=1}^{L}C^\text{L*}_l \notag\\
&= B t^*_{1} \log _2\left(1+\frac{\sum_{l=1}^Le^\text{o*}_l\Gamma_l}{t_1}\right)+\sum_{l=1}^{L}C^\text{L*}_l\notag\\
 &=B t^*_{1} \log _2\left(1+\frac{\sum_{l=1}^Le^\text{o*}_l\Gamma_l}{\sum_{l=1}^L\tau^*_l}\right)+\sum_{l=1}^{L}C^\text{L*}_l\notag\\
 &=B \sum_{l=1}^L \tau^*_{l} \log _2\left(1+\frac{e^\text{o*}_l\Gamma_l}{\tau^*_l}\right)+\sum_{l=1}^{L}C^\text{L*}_l.
 \end{align}

From (\ref{u3}), it is clear that the optimal solution of problem (2.2) is also a feasible solution of problem (1.2). Therefore, $O^{\mathrm{PD}*}_{\mathrm{TDMA}} \geq O^{\mathrm{PD}*}_{\mathrm{NOMA}}$. Combining this with  $O^{\mathrm{PD}*}_{\mathrm{TDMA}} \leq O^{\mathrm{PD}*}_{\mathrm{NOMA}}$, we conclude that $O^{\mathrm{PD}*}_{\mathrm{TDMA}} = O^{\mathrm{PD}*}_{\mathrm{NOMA}}$. \textcolor{black}{Moreover, following the same derivation process as in the proof of Theorem 3, it can be also proven that $O^{\mathrm{S}}_{\mathrm{TDMA}}  = O^{\mathrm{S}}_{\mathrm{NOMA}} $. Due to space limitations, we omit the detailed proof here.}
\end{IEEEproof}

Theorem 3 demonstrates that, for the WPT-MEC system with discrete PAs, the TDMA and NOMA schemes achieve identical computation performance in partially dynamic PA activation.  Combining Theorems 1 and 2 then yields the following relationship
 \begin{equation}\small
  O^{\mathrm{S}}_{\mathrm{NOMA}}=O^{\mathrm{S}}_{\mathrm{TDMA}} \leq  O^{\mathrm{PD}}_{\mathrm{NOMA}}=O^{\mathrm{PD}}_{\mathrm{TDMA}}=O^{\mathrm{FD}}_{\mathrm{NOMA}}\leq O^{\mathrm{FD}}_{\mathrm{TDMA}}.
 \end{equation}

Based on the above comparison, we observe that both the dynamic adjustment of PA activation strategies and the choice of access scheme significantly affect the computation performance of WPT-MEC system with discrete PAs. Among all PA activation configurations, fully dynamic PA activation under the TDMA scheme achieves the highest total computational bits, while partially dynamic PA activation consistently outperforms static PA activation under both schemes, underscoring the performance gains brought by dynamic PA activation. Furthermore, within each of static PA activation and partially dynamic PA activation, the TDMA and NOMA schemes achieve identical computation performance.
\section{Proposed Algorithm}
In this section, we focus on solving the optimization problems formulated in Section II.B. Due to the non-convex terms in the objective function with respect to the PA activation strategies, the intricate coupling among power allocation, transmit power, and PA activation, as well as the binary constraints (\ref{prob29}a) and (\ref{prob32}a), problems (P1.1)–(P2.3) are highly non-convex MINLP and thus are difficult to solve them  directly. To address this challenge, we consider (P1.2) as an illustrative example and propose a two-layer optimization algorithm. Specifically, by exploiting the hierarchical structure of the problem, we decompose (P1.2) into two subproblems, namely (i) an inner subproblem that optimizes the allocation of downlink and uplink phase durations $t_0$ and $\{\tau_l\}_{l=1}^L$, the transmit power $p_l$, and the computational frequency $f_l$ of the devices; and (ii) an outer subproblem that optimizes the PA activation strategies $\bm{\beta}^{\mathrm{D}}$ and $\{\bm{\beta}^{\mathrm{U}}_l\}_{l=1}^L$, which are respectively  formulated as
\begin{align} 
\label{prob34_1}
\left(\text{Inner}\right):  \max_{\substack{t_0,\tau_l,\\  \{p_l\}, \{f_l\}}} & ~R^{\mathrm{PD}}_{\mathrm{TDMA}}+\sum_{l=1}^{L}C^\text{L}_l\\
\text{s.t.}~~
&  \text{(\ref{prob29}\text{b})}, \text{(\ref{prob29}\text{c})},\text{(\ref{prob29}\text{d})}. \notag
\end{align}
\begin{align} 
\label{prob34_2}
\left(\text{Outer}\right):  \max_{\bm{\beta}^{\mathrm{D}},\bm{\beta}^{\mathrm{U}}} & ~R^{\mathrm{PD}}_{\mathrm{TDMA}}+\sum_{l=1}^{L}C^\text{L}_l\\
\text{s.t.}~~
& (\text{\ref{prob30}a}). \notag
\end{align}

 For the inner subproblem (\ref{prob34_1}), we derive closed-form solutions for $t_0$,  $\{\tau_l\}_{l=1}^L$, $p_l$, and $f_l$ using the KKT conditions. For the outer subproblem (\ref{prob34_2}), we propose a cross-entropy-based algorithm to learn the probability distribution of PA activation strategies. At each iteration, we randomly sample $\bm{\beta}^{\mathrm{D}}$ and $\bm{\beta}^{\mathrm{U}}$  from the current distribution, and the objective value of the inner subproblem is evaluated for each sample. The distribution is then updated accordingly, gradually guiding the search toward the optimal solution. Finally, we demonstrate that the proposed two-layer algorithm can also be extended to other PA activation configurations.

\subsection{Closed-Form Solution for Solving the Inner Subproblem}
To  reduce computational complexity of solving the inner subproblem (\ref{prob34_1}), we derive a closed-form solution by applying the KKT conditions. Specifically, we adopt the same definitions as those used in the proof of Theorem 3, i.e., $\Gamma_l=\tfrac{G^{\mathrm{U}}_l(\bm{\beta}^{\mathrm{U}})}{||\bm{\beta}^{\mathrm{U}}||_0\sigma_B^2}$, $e^\text{o}_l = p_l\tau_l$, and $z_l \triangleq \tfrac{e^\text{o}_l\Gamma_l}{\tau_l}$.  With these definitions, (\ref{prob34_1}) can be reformulated as
\begin{align} 
\label{prob35}
  \max_{\substack{t_0,\{\tau_l\},\\  \{e_l^{\mathrm{o}}\}, \{f_l\}}} & ~B \sum_{l=1}^L \tau_{l} \log _2\left(1+\frac{e^\text{o}_l\Gamma_l}{\tau_l}\right)+\sum_{l=1}^{L}C^\text{L}_l\\
\text{s.t.}~~
&  \text{(\ref{prob29}\text{b})}, \text{(\ref{prob29}\text{c})},\text{(\ref{prob29}\text{d})}. \notag
\end{align}

Its Lagrangian function can be written as\footnote{We consider only the coupling constraints in the Lagrangian function, while the simple non-negativity constraints, e.g., (13d), are handled by projecting the obtained solution onto the feasible set.}

\begin{align}
\mathcal{L}=&B  \sum_{l=1}^{L} \tau_{l} \log _2\left(1+\frac{e^\text{o}_l\Gamma_l}{\tau_l}\right)+\sum_{l=1}^{L}\alpha_l\left(E_l-e^{\text{L}}_l-e^\text{o}_l\right)
\notag\\&\quad+\lambda\left(T-t_0-\sum_{l=1}^L \tau_{l}\right),
\end{align}
where $\{\alpha_l\}_{l=1}^L$ and $\lambda$ denote the  Lagrange multipliers associated with the constraint (\ref{prob29}\text{b}) and (\ref{prob29}\text{c}), respectively. The  KKT conditions can be expressed as

\begin{align}
&\frac{\partial \mathcal{L}}{\partial \tau_{l}} =0,~\frac{\partial \mathcal{L}}{\partial t_{0}}=0, ~\frac{\partial \mathcal{L}}{\partial f_{l}} =0, ~~\frac{\partial \mathcal{L}}{\partial e^{\text{0}}_{l}} =0,\alpha_l\geq 0, \forall l,\lambda\geq 0,
\notag\\&\alpha_l\left(E_l-e^{\text{L}}_l-e^\text{o}_l\right)=0,\forall l,~~ \lambda\left(T-t_0-\sum_{l=1}^L \tau_{l}\right)=0,\notag\\
&E_l-e^{\text{L}}_l-e^\text{o}_l\geq 0,  \forall l,~~ T-t_0-\sum_{l=1}^L \tau_{l}\geq 0.
\end{align}

We define devices performing uplink offloading as active users,  denoted by $\mathcal{A}=\{\hat{l}\mid e^\text{o}_{\hat{l}}>0,\tau_{\hat{l}}>0\}$, while devices performing only local computation are considered inactive. We initially assume that all devices belong to the active set $\mathcal{A}$.  From $\tfrac{\partial \mathcal{L}}{\partial \tau_{l}} =0$, we obtain
\begin{equation}
\label{v1}
\lambda =  B \left( \log_2(1+z_l) - \frac{z_l}{(1+z_l)\ln 2} \right), \forall l\in \mathcal{A}.
\end{equation}

Since $\lambda$ is identical for $ \forall l\in \mathcal{A}$, it follows that $z=z_l, \forall l\in \mathcal{A}$. From $\tfrac{\partial \mathcal{L}}{\partial f_l} =0$, we  obtain
\begin{equation}
\label{v8}
f_l^{*}=
\begin{cases}
\sqrt{\frac{1}{3 \kappa I_c\alpha_l}}, & \forall\, l \in \mathcal{A}, \\[2ex]
\sqrt[3]{\frac{t_0\Upsilon_l}{\kappa T}}, & \forall\, l \notin \mathcal{A}. 
\end{cases}
\end{equation}

From   $\tfrac{\partial \mathcal{L}}{\partial e^{\text{o}}_l} =0$, we  obtain

\begin{equation}
\label{v3}
\alpha_l =
\begin{cases}
\dfrac{B}{\ln 2}\dfrac{\Gamma_l}{1+z}, & \forall\, l \in \mathcal{A}, \\[2ex]
\dfrac{1}{3\kappa I_c}\left(\dfrac{T\kappa}{t_0\Upsilon_l}\right)^{\frac{2}{3}}, & \forall\, l \notin \mathcal{A},
\end{cases}
\end{equation}

Moreover, from $\tfrac{\partial \mathcal{L}}{\partial t_0} =0$ and by defining $\Upsilon_l=\gamma \tfrac{P_b}{||\bm{\beta}^{\mathrm{D}}||_0} G^{\mathrm{D}}l(\bm{\beta}^{\mathrm{D}}), \forall l$, we have
\begin{equation}
\label{v2}
\lambda =  \sum_{l=1}^{L}\Upsilon_l, \forall l.
\end{equation}

By substituting (\ref{v1}) and (\ref{v3}) into (\ref{v2}), we obtain
\begin{equation}
\label{v4}
 \ln(1+z) - \frac{z}{(1+z)} =  \frac{\sum_{l\in\mathcal{A}}\Upsilon_l\Gamma_l}{1+z}+\sum_{l\notin\mathcal{A}}\frac{\ln2\Upsilon^{\frac{1}{3}}_l}{3B\kappa I_c}\left(\dfrac{T\kappa}{t_0}\right)^{\frac{2}{3}},
\end{equation}
which constitutes a coupled equation in the variables $t_0$ and $z$. Since each device fully utilizes its harvested energy, we have
\begin{equation}
\label{v5}
e_l^{\mathrm{o}*} = \left[t_0\Upsilon_l -T\kappa f^{*3}_l\right]_+
\end{equation}
where $\left[\,\cdot\,\right]_+$ denotes the operator $\max\{\,\cdot\,,~0\}$. Moreover, as the devices also make full use of both downlink and uplink phase durations, it follows that $T-t^*_0-\sum_{l=1}^L \tau^*_{l}=0$. Combining this with $z^* = \tfrac{e^\text{o*}_l\Gamma_l}{\tau^*_l}$,  we obtain
\begin{equation}
\label{v6}
t^*_0 = T-\frac{\sum_{l\in\mathcal{A}}\Gamma_le_l^{\mathrm{o}*}}{z^*}.
\end{equation}

Substituting (\ref{v5}) into (\ref{v6}) yields the closed-form solution for $t_0$:
\begin{equation}
\label{v7}
t_0^{*}=\frac{z^{*} T+T \kappa \sum_{l\in\mathcal{A}} \Gamma_l\left(f_l^{*}\right)^3}{z^{*}+\sum_{l\in\mathcal{A}}\Upsilon_l \Gamma_l}.
\end{equation}

Since $f_l^{*}$ is a term that depends on $z^*$, the unique root $z^*$ can be efficiently obtained by jointly solving (\ref{v4}) and (\ref{v7}). Once $z^*$ is determined, the corresponding $\alpha_l$ and $t_0^{*}$ can be computed from (\ref{v3}) and (\ref{v7}), respectively. Subsequently, $f_l^{*}$ and $e_l^{\mathrm{o*}}$ are derived from (\ref{v8}) and (\ref{v5}), while $\tau^*_l$ and $p^*_l$ can then be obtained by
\begin{equation}
\label{v9}
\tau^*_l = \frac{e^{\mathrm{o}*}_l\Gamma_l}{z^*},~~ p^*_l = \frac{e^\text{o*}_l}{\tau^*_l}.
\end{equation}

 Finally, we use (\ref{v5}) to check whether the resulting active device set $\mathcal{A}$ matches the initially assumed one: a device is considered inactive if $e_l^{\mathrm{o*}} = 0$. If the two sets coincide, the optimal solutions $\{f_l^{*}, t^*_0, \tau^*_l, p^*_l\}$ are obtained; otherwise, $\mathcal{A}$ is updated and the procedure is repeated. The iteration continues until the active device set remains unchanged. The detailed steps for solving the inner subproblem (\ref{prob35}) are summarized in Algorithm~1.

\renewcommand{\algorithmicrequire}{\textbf{Input:}}
\renewcommand{\algorithmicensure}{\textbf{Output:}}
\begin{algorithm}[t] 
	\caption{ Algorithm for solving the inner subproblem  (\ref{prob34_1})}
	\begin{algorithmic}[1] 
		\State \textbf{Input:} Downlink and uplink PA positions $ \bm{\beta}^{\mathrm{D}}, \bm{\beta}^{\mathrm{U}}$, and initial active device set $\mathcal{A}$.
		\Repeat
		\State Solve (\ref{v4}) and (\ref{v7}) jointly to obtain $z^*$.
		\State Compute the downlink phase duration $t^*_0$ from (\ref{v7}).
		\State Compute the Lagrange multipliers $\alpha_l$ from (\ref{v3}).
		\State Compute $e_l^{\mathrm{o}*}$ from (\ref{v5}).
		\State Compute the uplink sub-slot $\tau^*_l$ via (\ref{v9}).
		\State Compute the transmit power $p^*_l$ via (\ref{v9}).
		\State Update the active device set $\mathcal{A}$ based on (\ref{v5}).
		\Until {$\mathcal{A}$ remains unchanged.}
		\State \textbf{Output:} Optimal downlink phase duration $t^*_0$, uplink sub-slots $\tau^*_l$, transmit powers $p^*_l$, and computational frequencies $f^*_l$ for all devices.
	\end{algorithmic}
\end{algorithm}

\subsection{Cross-Entropy Algorithm for the Outer Subproblem}
In this subsection, we propose a cross-entropy-based algorithm to address the outer integer programming problem (\ref{prob34_2}). The cross-entropy method is a general probabilistic optimization framework that has been widely applied in machine learning and is particularly effective for complex non-convex problems \cite{CE2,CE3}. Its core idea is to sample candidate solutions from a probability distribution, evaluate their performance, and iteratively update the distribution parameters so that the search gradually concentrates on the most promising regions of the solution space. Compared with convex relaxation methods or heuristic algorithms that often suffer from performance degradation, the cross-entropy method exploits large-scale sampling to approximate the optimal distribution of decision variables, thereby offering the potential to approach the global optimum. The main steps of the algorithm include sample generation, elite  sample selection, and distribution update.
\subsubsection{Sample Generation}
Sample generation is performed by drawing random candidate solutions from a probability distribution defined over the optimization variables. The choice of distribution function depends on the type of variables. Since the PA activation variables in problem (\ref{prob34_2}) are binary, we model them using a Bernoulli distribution. Specifically, let $\bm{\upsilon}^{\mathrm{D}}=\{\upsilon^{\mathrm{D}}_{l}\}_{l=1}^L$ denote the Bernoulli parameters for the downlink PA activation variables, where $\upsilon^{\mathrm{D}}_{l}$ represents the probability that $\beta^{\mathrm{D}}_l=1$. The probability distribution of $\bm{\beta}^{\mathrm{D}}$ is then given by
\begin{equation}
\label{p2}
\mathscr{P}^{\mathrm{D}}(\bm{\beta}^{\mathrm{D}}; \bm{\upsilon}^{\mathrm{D}}) \triangleq \prod_{l=1}^L (\upsilon^{\mathrm{D}}_{l})^{\beta^{\mathrm{D}}_l}\left(1-\upsilon^{\mathrm{D}}_{l}\right)^{1-\beta^{\mathrm{D}}_l}.
\end{equation}

Similarly, the probability distribution function of the uplink PA activation strategy  $\bm{\beta}^{\mathrm{U}}$, parameterized by $\bm{\upsilon}^{\mathrm{U}}=\{\upsilon^{\mathrm{U}}_{l}\}_{\forall l}$, is modeled as
\begin{equation}
\label{p3}
\mathscr{P}^{\mathrm{U}}(\bm{\beta}^{\mathrm{U}}; \bm{\upsilon}^{\mathrm{U}}) \triangleq \prod_{l=1}^L (\upsilon^{\mathrm{U}}_{l})^{\beta^{\mathrm{U}}_l}\left(1-\upsilon^{\mathrm{U}}_{l}\right)^{1-\beta^{\mathrm{U}}_l}.
\end{equation}

Since the uplink and downlink PA activation strategies are independent, their joint probability distribution can be expressed as
\begin{equation}
\label{p1}
\mathscr{Q}(\bm{\beta}_\mathrm{J}; \bm{\upsilon}_\mathrm{J}) \triangleq \mathscr{P}^{\mathrm{D}}(\bm{\beta}^{\mathrm{D}}; \bm{\upsilon}^{\mathrm{D}}) \times \mathscr{P}^{\mathrm{U}}(\bm{\beta}^{\mathrm{U}}; \bm{\upsilon}^{\mathrm{U}}),
\end{equation}
where $\bm{\beta}_\mathrm{J}=\{\bm{\beta}^{\mathrm{U}},\bm{\beta}^{\mathrm{D}}\}$ and $\bm{\upsilon}_\mathrm{J}=\{\bm{\upsilon}^{\mathrm{U}},\bm{\upsilon}^{\mathrm{D}}\}$. We initialize $\upsilon_l^{\mathrm{D}}=\upsilon_l^{\mathrm{U}}=0.5$ for all $l$, and randomly generate $S$ solution samples $\{\bm{\beta}^s_\mathrm{J}\}_{s=1}^S$ based on the joint probability distribution in (\ref{p1}).
\subsubsection{Elite  Selection}
For each generated sample, we solve the corresponding inner optimization problem (\ref{prob35}) using Algorithm 1 and record its objective value. Since Algorithm 1 exploits closed-form solutions, the optimization is computationally lightweight even when performed across multiple evaluations. The obtained objective values are then sorted in descending order, and the top $S_\mathrm{E}$ samples are selected as elite samples $\{\bm{\beta}^{[1]}_\mathrm{J},...,\bm{\beta}^{[S_\mathrm{E}]}_\mathrm{J}\}$, which are subsequently used to update the Bernoulli parameters.
\subsubsection{Distribution Update}
After selecting the elite samples, we update the Bernoulli parameters $\bm{\upsilon}_\mathrm{J}$ of the distribution (\ref{p1}). Specifically, the goal is to minimize the cross-entropy (or Kullback–Leibler divergence) between the current distribution and the empirical distribution of the elite samples. According to \cite{CE3}, the cross-entropy minimization problem can be reformulated as the following optimization problem
\begin{align} 
\label{CE4}
\min_{\bm{\upsilon}_\mathrm{J}} &  ~~-\frac{1}{{S}}\sum_{s=1}^{{S}_\mathrm{E}} \ln \mathscr{Q}(\bm{\beta}^{[s]}_\mathrm{J}; \bm{\upsilon}_\mathrm{J}).
\end{align}

By substituting (\ref{p2}) and (\ref{p3}) into (\ref{CE4}) and setting the first-order derivatives of the objective function with respect to $\upsilon^{\mathrm{D}}_{l}$ and $\upsilon^{\mathrm{U}}_{l}$ to zero, the optimal parameter updates are obtained as
\begin{equation} \label{CE6}
\upsilon^{\mathrm{D}*}_{l} = \frac{1}{{S}_\mathrm{E}}\sum_{s=1}^{{S}_\mathrm{E}} (\beta^{\mathrm{D}}_{l})^{{[s]}},~~~\upsilon^{\mathrm{U}*}_{l} = \frac{1}{{S}_\mathrm{E}}\sum_{s=1}^{{S}_\mathrm{E}} (\beta^{\mathrm{U}}_{l})^{{[s]}}.
\end{equation}

\renewcommand{\algorithmicrequire}{\textbf{Input:}}
\renewcommand{\algorithmicensure}{\textbf{Output:}}
\begin{algorithm}[t] 
	\caption{Algorithm for the outer subproblem  (\ref{prob34_2})}
	\begin{algorithmic}[1] 
		\State {Initialize the iteration number $i = 0$, and set $\upsilon_l^{\mathrm{D}(0)} = \upsilon_l^{\mathrm{U}(0)} = 0.5$ for all $l$.}
		\Repeat
		\State {Generate $S$ solution samples according to the distribution (\ref{p1}).}
		\State {For each sample, compute the objective value of problem (\ref{prob34_1}) by applying \textbf{Algorithm 1}.}
		\State {Sort the $S$ samples in descending order of objective values.}
		\State {Select the top ${S}_\mathrm{E}$ samples as elite samples.}
		\State {Compute $\hat{\bm{\upsilon}}_\mathrm{J}$ based on (\ref{CE6}).}
		\State {Update $\bm{\upsilon}_\mathrm{J}^{(i+1)}$ based on (\ref{CE7}).}	  	    
		\State {Update $i = i + 1$.}
		\Until {the objective value converges.}	
	\end{algorithmic}
\end{algorithm}

The update in (\ref{CE6}) adjusts the Bernoulli parameters so that subsequent samples are more likely to fall within high-quality regions of the solution space. To further improve convergence stability, we adopt a smoothing mechanism commonly used in reinforcement learning, where a  smoothing parameter $\zeta$ is introduced to form a convex combination of the old and newly updated parameters. Denoting the Bernoulli parameters at the $(i)$-th iteration as $\bm{\upsilon}^{(i)}_\mathrm{J}$ and those obtained from (\ref{CE6}) as $\hat{\bm{\upsilon}}_\mathrm{J}$, the smoothed update is given by
\begin{equation} \label{CE7}
\bm{\upsilon}_\mathrm{J}^{(i)} =(1-\zeta)\bm{\upsilon}_\mathrm{J}^{(i)}+\zeta\hat{\bm{\upsilon}}_\mathrm{J}^{(i)}.
\end{equation}

After updating the Bernoulli parameters, we regenerate candidate samples, select the elite set, and update the distribution accordingly. This procedure is iteratively repeated until the objective value converges. For clarity, the detailed steps are summarized in Algorithm 2.

\subsection{Extension to Other PA activation configurations}
In this subsection, we demonstrate that the proposed two-layer optimization algorithm  can also be extended to other PA activation configurations. Specifically, when Algorithm 2 considers only the probability distribution of a single PA activation strategy, the proposed algorithm can be applied to solve the optimization problem (P1.1) for static PA activation under the TDMA scheme. Similarly, when the algorithm accounts for the probability distribution of 
${(L+1)}$ PA activation strategies ($L$ uplink sub-slots and downlink), the overall algorithm can be applied to solve the optimization problem (P1.3) for fully dynamic PA activation. After obtaining the solutions under the TDMA scheme, as detailed in Section III, they can be readily reconstructed to address the corresponding optimization problems under the NOMA scheme.
\subsection{Complexity and Convergence Analysis}
The computational complexity of the proposed two-layer algorithm primarily arises from the cross-entropy-based algorithm used to solve the outer optimization problem, which requires multiple calculations of the inner optimization problem's objective value. Specifically, the inner optimization problem can be solved in  closed-form with a computational complexity of $\mathcal{O}(L)$. Considering the multiple sampling operations in the outer layer, the overall complexity of the algorithm is $\mathcal{O}(I_{it} S L)$, where $I_{it}$ represents the number of iterations, and $S$ is the number of samples generated in each iteration.

\textcolor{black}{Regarding the convergence of Algorithm 2, it employs the cross-entropy method, which converges when the number of samples is sufficiently large and an optimal importance sampling distribution is available \cite{CEconv}. In our case, the PA activation variables define a finite, discrete solution space, ensuring the existence of such a distribution. To approximate this distribution, we use a product of independent Bernoulli distributions, which effectively model binary variables and facilitate efficient parameter updates. Therefore, Algorithm 2 is guaranteed to converge under these conditions.}
\textcolor{black}{1}

\section{Simulation Results}

 \begin{table}[t]
	\renewcommand{\arraystretch}{1.25}
	\caption{SIMULATION PARAMETERS.}
	\label{table_example2}
	\centering
	\begin{tabular}{l l}
		\hline
		
		\bfseries Parameters &  \multicolumn{1}{c}{\bfseries Value}\\ 
		\hline
		Number of devices $L$ & 3\\
		Number of Discrete PAs $N$  & 40\\
		Length of transmission frame $T$  & 1 s  \\
		\tabincell{l}{BS' transmit power $P_{b}$} & 43 dBm\\
		Noise power at the BS & -120 dBm\\
		Computation intensity $Ic$ & 200 Cycles/bit \\
		Carrier frequency $f_c$ & 28 GHz\\
		Signal Bandwidth $B$ & 50 MHz\\
		Height of waveguide $d$ &4~m\\
		Effective refractive index of the dielectric waveguide $n_e$ & 1.4  \\
		The  power coefficient of devices $\kappa$ & $10^{-28}$ \\
		The energy conversion efficiency of devices $\gamma$ &0.8\\
		Number of samples $S$ &500\\
			Number of elite samples  $S_\mathrm{E}$ &50\\
			 Smoothing parameter $\zeta$ &0.9\\
		\hline
		
	\end{tabular}
\end{table}

In this section, we present simulation results to demonstrate the performance gains of discrete PAs in WPT-MEC systems, as well as the effectiveness of the proposed two-layer optimization algorithm. Furthermore, we provide a numerical analysis of how dynamically adjusting PA activation strategies affects the system’s computation performance.

\subsection{System Setup and Benchmarking Schemes}
For the system setup, we consider  $L=3$ devices randomly distributed within a rectangular region of $30~\text{m}\times10~\text{m}$. The waveguide connected to the BS is placed at a height of $d=4$ m. The carrier frequency and bandwidth of the signals are set as $f_c=28$ GHz and $B=50$ MHz, respectively. The BS transmit power is $P_b=43$ dBm, and the noise power is ${\sigma_B^2}=-120$ dBm. In addition, the effective refractive index of the dielectric waveguide is set as $n_e=1.4$ \cite{UPASS1}, and the computation density is assumed to be $I_c=200$ cycles/bit. Other key parameters are summarized in Table 	\ref{table_example2}. Unless otherwise specified, the ``discrete PAs” scheme refers to the numerical results of \textbf{partially dynamic PA activation} under the TDMA framework. To evaluate the performance gains of discrete PAs, we compare it against the following five baseline schemes:
\begin{itemize}
\item \textbf{Full PA activation (Full PA) scheme}:	In this scheme, all PAs are activated for signal transmission during both the downlink WPT phase and the uplink offloading phase. The optimization of the remaining variables follows the same procedure as in the discrete PA scheme.  
\item \textcolor{black}{\textbf{Conventional antenna (CA) scheme}:} A uniform linear array  with $N$ antennas is employed for both downlink WPT and uplink signal reception, located at $(0,0,d)$ with an antenna spacing of half a wavelength.
\item \textbf{TDMA with fixed time-slot allocation (F-TDMA) scheme}: The entire transmission frame is evenly divided into $L+1$ sub-slots, where the first sub-slot is dedicated to downlink WPT and the remaining sub-slots are allocated to different devices for uplink offloading.
\item \textbf{Full offloading (F-Offload) scheme}: All harvested energy is allocated to task offloading, while the timeslot allocation and PA activation strategy optimization remain the same as in the discrete PA scheme.  
\item \textbf{Full local computing (F-Local) scheme }: All harvested energy is used exclusively for local computation, with CPU frequency selection and PA activation strategy optimization performed in the same way as in the discrete PA scheme.
\end{itemize}

 \begin{figure}[!t]
	\centering
	\includegraphics[width=3.1in]{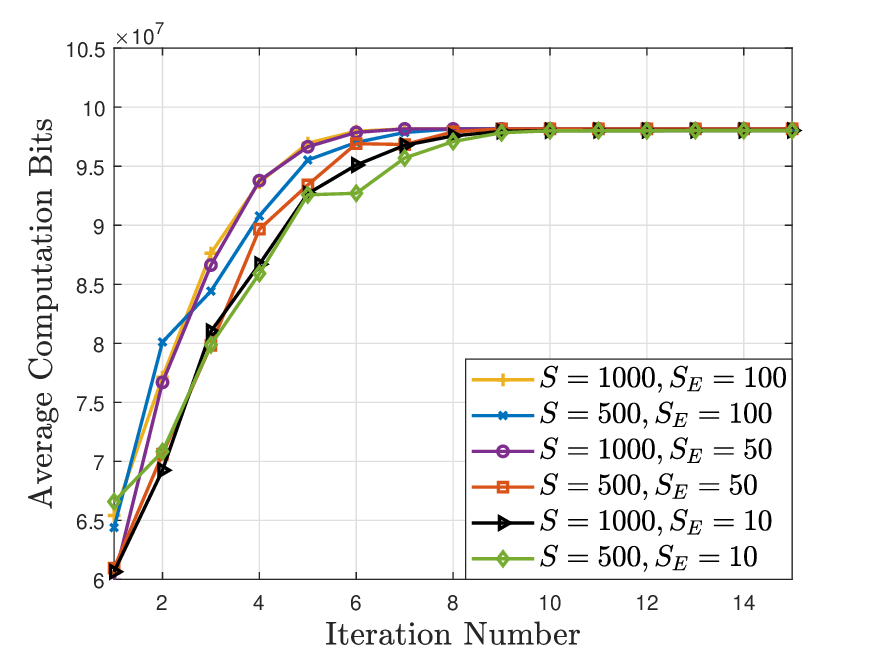}
	\caption{The convergence performance of proposed algorithm.}
	\label{fig:c}
\end{figure}

\subsection{Convergence  Performance}

We first demonstrate the convergence behavior of the proposed two-layer optimization algorithm. As shown in Fig.~\ref{fig:c}, under different numbers of samples $S$ and elite samples $S_\mathrm{E}$, the algorithm consistently converges to the same objective value within 10 iterations. Moreover, increasing $S$ provides a more accurate estimation of the objective distribution, thereby reducing variance and enabling more reliable parameter updates. Similarly, a larger elite set enhances the robustness of the distribution update by alleviating the risk of overfitting to a small number of candidate solutions. Consequently, both larger $S$ and $S_\mathrm{E}$ accelerate convergence while preserving stability.

\subsection{Performance  Comparison}
 \begin{figure}[!t]
	\centering
	\includegraphics[width=3.5in]{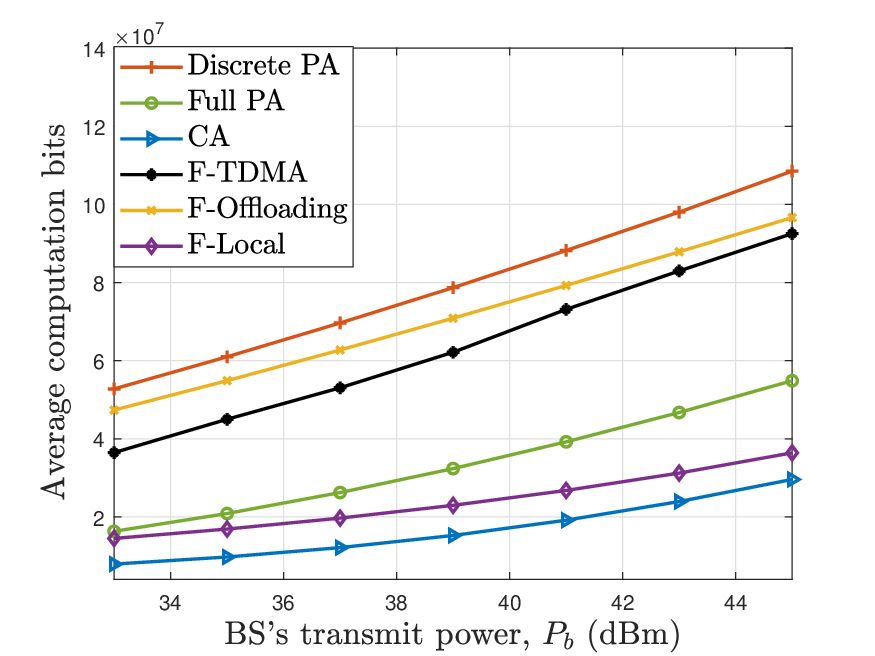}
	\caption{Average computational bits versus BS's transmit power $P_b$.}
	\label{fig:0}
\end{figure}
In Fig.~\ref{fig:0}, \textcolor{black}{we present the average computational bits of $L$ devices $\tfrac{O^{\mathrm{PD}*}_{\mathrm{TDMA}} }{L}$ under different schemes  with respect to the BS transmit power $P_b$.} It is observed that increasing $P_b$ enables devices to harvest more energy, thereby improving the average computation performance across all schemes. Notably, the discrete PA scheme consistently outperforms the baselines. In particular, compared with the CA scheme, the dynamic activation of PAs effectively reduces the path loss between the transmission points and the devices, which in turn enhances both WPT efficiency and uplink offloading rate. Relative to the F-TDMA scheme, the more flexible time-slot allocation offered by the discrete PA scheme provides a better trade-off between downlink energy transfer and uplink task offloading, thus maximizing the utilization of time resources. Moreover, the discrete PA scheme achieves superior performance compared with the F-Offloading and F-Local schemes. This is because exclusive reliance on task offloading is constrained by the limited uplink transmission duration, whereas purely local computing suffers from low computation efficiency; jointly exploiting both modes enables more effective utilization of the harvested energy. Finally, it is worth highlighting that the discrete PA scheme surpasses the Full PA scheme, showing that activating all PAs is not always advantageous. Owing to phase superposition effects, indiscriminate PA activation can even degrade performance, whereas selective PA activation ensures constructive alignment, thereby enhancing both energy transfer and uplink offloading efficiency.
 
\begin{figure}[!t]
	\centering
	\includegraphics[width=3.51in]{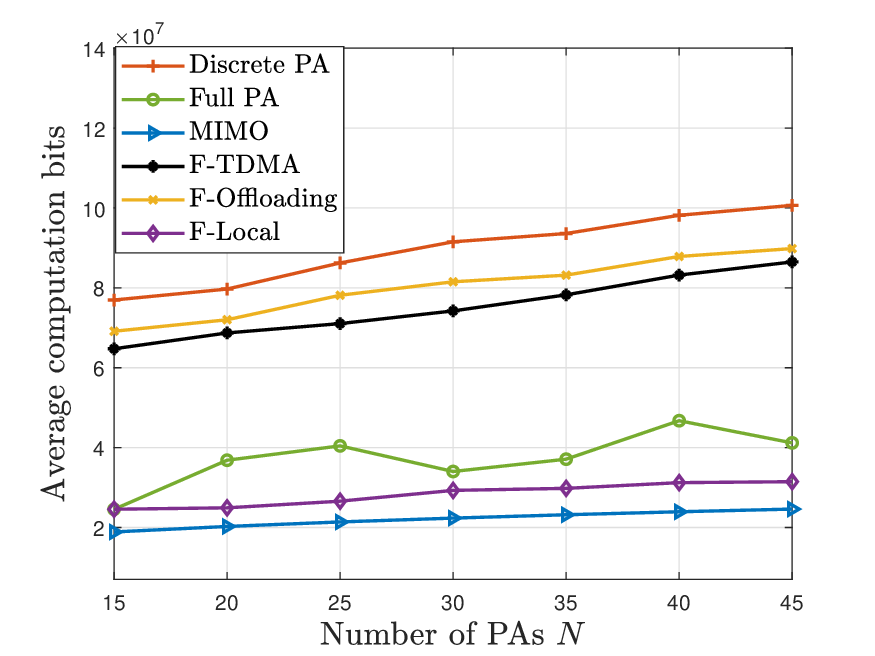}
	\caption{Average computational bits versus the number of antennas $N$.}
	\label{fig:1}
\end{figure}

In Fig.~\ref{fig:1}, we show the impact of the number of antennas $N$ on the average computational bits under different schemes. It can be observed that the proposed discrete PA scheme consistently achieves the highest average computational bits across all antenna configurations. As $N$ increases, the number of potential PA activation positions also grows, enabling the BS to more precisely adjust the activation strategy to enhance both WPT and task offloading performance. Moreover, the performance gain achieved by increasing $N$ is more pronounced in the discrete PA scheme than in the CA scheme. This is because the discrete PA scheme benefits from reduced path loss with more antennas, while the gain in the CA scheme mainly comes from beam focusing rather than path loss reduction. We also observe that the computation performance of the Full PA scheme does not increase linearly with the number of antennas. This is due to the inability of the full-activation strategy to ensure phase coherence among all PAs as $N$ varies, which leads to performance variability.

\begin{figure}[!t]
	\centering
	\includegraphics[width=3.5in]{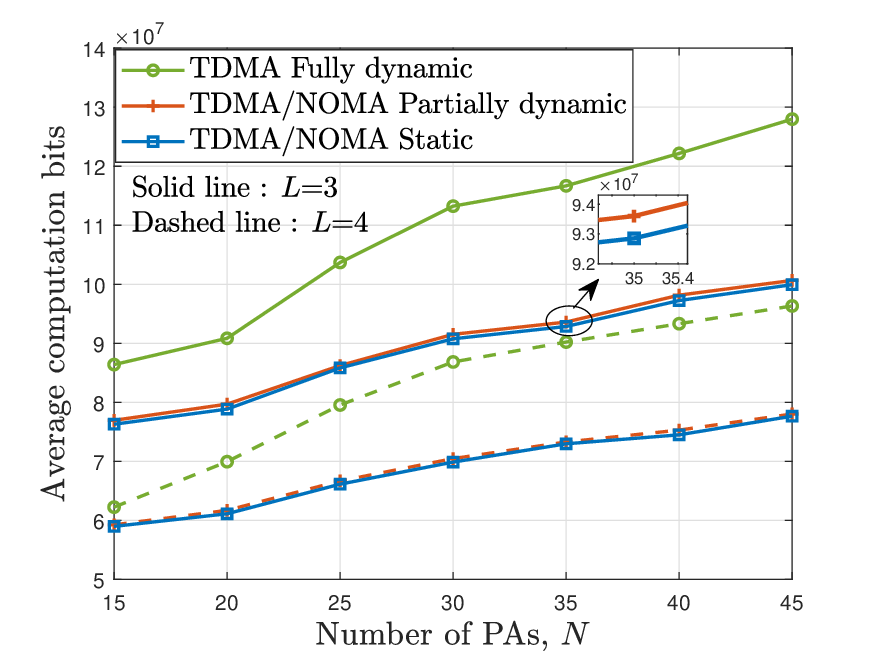}
	\caption{Average computational bits under different PA activation configurations.}
	\label{fig:2}
\end{figure}

To evaluate the impact of dynamic PA activation strategies on system performance, Fig.~\ref{fig:2} presents the average computational bits under different PA activation configurations. We observe that \textbf{fully dynamic PA activation} of TDMA achieves the highest performance among all PA activation configurations by flexibly activating different PAs in both the downlink phase duration and the uplink sub-slots, demonstrating the substantial gains achievable through dynamic PA activation. \textbf{Partially dynamic PA activation} under both TDMA and NOMA also shows moderate improvement over \textbf{static PA activation}, as PA activation can be varied between downlink and uplink phases. These numerical results align with the theoretical analysis in Section III, confirming its validity. In addition, the average computational bits increase with the number of antennas $N$ in all PA activation configurations, with \textbf{fully dynamic PA activation} under TDMA  benefiting the most. This indicates that  more discrete PAs are especially advantageous when dynamic activation is allowed. Fig.~\ref{fig:2} further illustrates the effect of the number of devices $L$. As $L$ grows, the average computational bits decrease under all PA activation configurations. This degradation arises from smaller uplink sub-slot allocations per device in TDMA and increased interference in NOMA, both of which reduce the offloaded computational bits.

\begin{figure}[!t]
	\centering
	\includegraphics[width=3.5in]{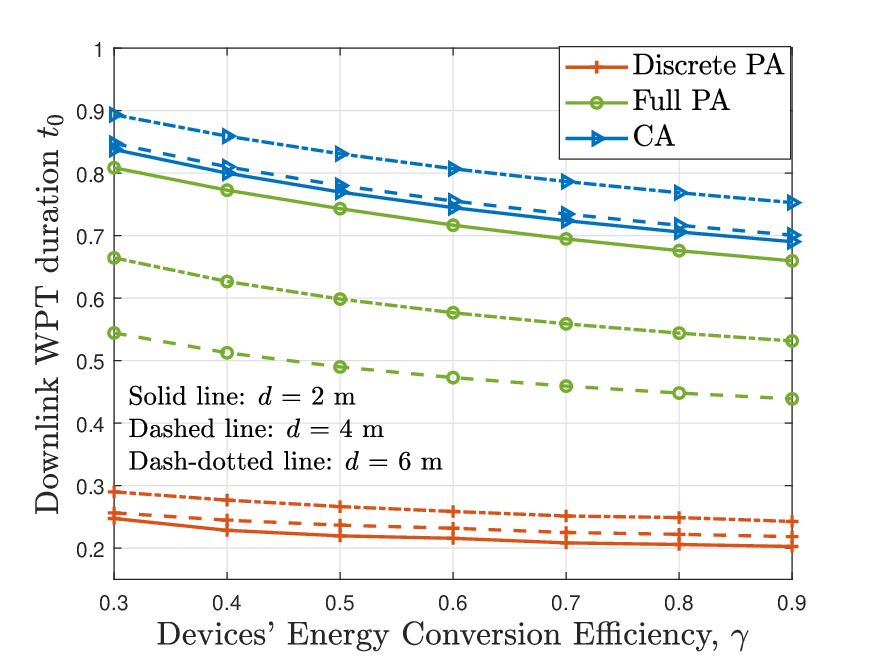}
	\caption{Downlink WPT duration $t_0$ versus devices' energy conversion efficiency $\gamma$.}
	\label{fig:3}
\end{figure}

To further illustrate the advantages of PAs, Fig.~\ref{fig:3} shows the variation of downlink WPT duration $t_0$  with respect to the devices' energy conversion efficiency $\gamma$ under three schemes: discrete PA, Full PA, and CA. As $\gamma$ increases, the WPT capability of devices improves, resulting in a shorter required $t_0$. Notably, the discrete PA scheme achieves the shortest downlink WPT duration among the three schemes. This advantage stems from its ability to reduce path loss through selective PA activation, thereby enhancing power transfer efficiency, delivering more energy within a shorter duration, and leaving additional uplink time for task offloading. We further examine the impact of waveguide height $d$ on $t_0$. For both the discrete PA and CA schemes, a larger $d$ increases path loss, thus requiring a longer $t_0$ to supply sufficient energy to the devices. In contrast, the Full PA scheme exhibits non-monotonic behavior, with the minimum $t_0$ observed at $d=4$ m. This is because variations in $d$ change the phase superposition characteristics of the Full PA scheme, so the downlink phase duration does not necessarily decrease monotonically with $d$.

\begin{figure}[!t]
	\centering
	\includegraphics[width=3.5in]{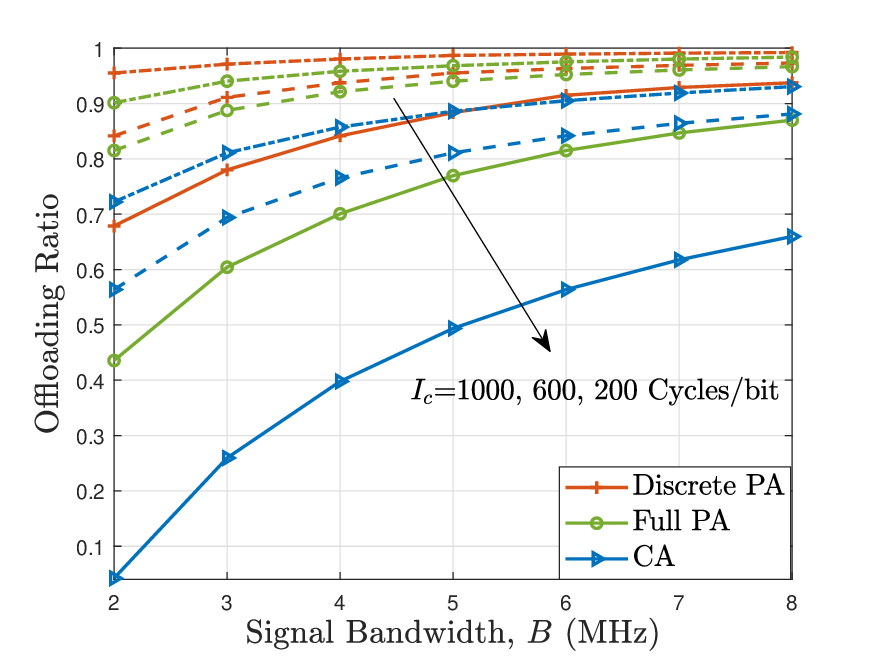}
	\caption{Offloading ratio versus the  signal bandwidth $B$.}
	\label{fig:4}
\end{figure}
Fig.~\ref{fig:4} illustrates the ratio of offloaded bits to the total computational bits under different signal bandwidths $B$. As $B$ increases, the offloading ratio rises across all three schemes, since a larger bandwidth enables higher offloading rates and makes devices more inclined to offload their tasks to the MEC server. The discrete PA scheme consistently achieves a higher offloading ratio than the other two schemes, owing to its lower path loss and greater offloading efficiency. We also examine the effect of computation intensity $I_c$ on the offloading ratio. As $I_c$ decreases, the burden of local computation is reduced, leading to a lower offloading ratio across all schemes. Moreover, reducing $I_c$ have the strongest impact on the CA scheme and the weakest impact on the discrete PA scheme. This demonstrates the robustness of the discrete PA design, where enhanced uplink channel gains ensure that offloading remains dominant and the system is less sensitive to changes in local computation requirements.
\begin{figure}[!t]
	\centering
	\includegraphics[width=3.5in]{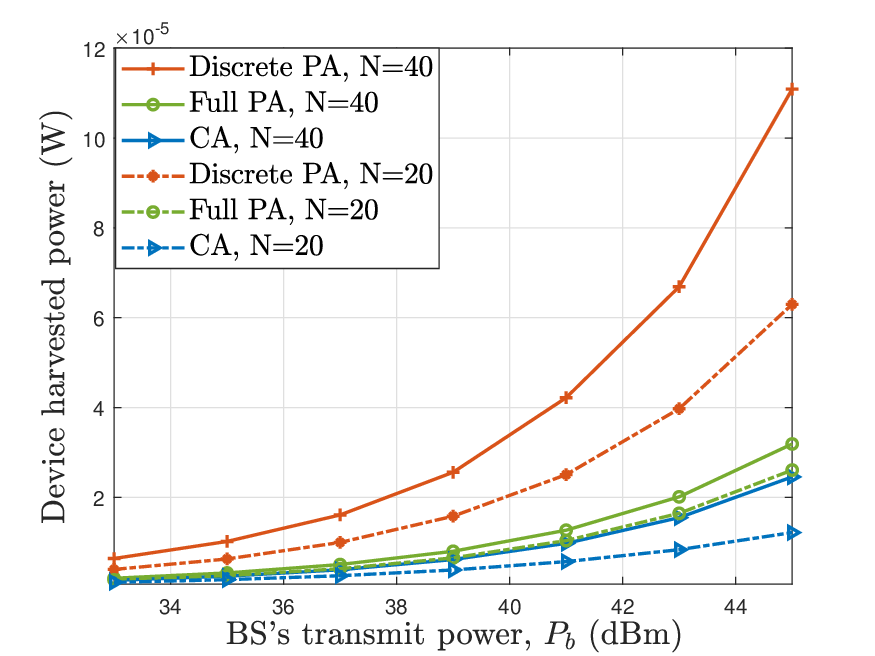}
	\caption{Device harvested power versus BS's transmit power $P_b$.}
	\label{fig:5}
\end{figure}

To evaluate the WPT gains of PAs, Fig.~\ref{fig:5} presents the average harvested power of devices $\frac{\sum^L_{l=1}{E}_{l}}{Lt_0}$ under different schemes versus the BS transmit power $P_b$. As $P_b$ increases, all schemes exhibit a corresponding growth in harvested power. Benefiting from significantly reduced path loss and enhanced phase alignment, the discrete PA scheme achieves significantly higher harvested power than both the CA and Full PA schemes, with the performance gap widening as $P_b$ increases. Furthermore, we examine the impact of the number of antennas $N$ on WPT. As $N$ grows, the harvested power improves across all schemes, with the discrete PA scheme showing the most pronounced gain. This observation highlights the advantage of selective PA activation, especially in large-scale antenna deployments.

\section{Conclusion}
In this paper, we investigated a discrete PA-assisted WPT-MEC framework, where devices harvest energy from PA-emitted RF signals and operate in a partial offloading mode. Both TDMA and NOMA were considered as uplink access schemes, and three levels of PA activation granularity were analyzed. For each PA activation configuration, we formulated a joint optimization problem to maximize the total computational bits by optimizing timeslot allocation, device transmit power, computational frequency, and PA activation strategies. We developed a theoretical comparison framework that revealed the performance differences across access schemes and activation strategies, and showed that TDMA achieved superior performance under more flexible activation. To solve the formulated problems, we designed a two-layer algorithm in which the inner problem exploited KKT conditions to derive closed-form solutions, while the outer problem applied a cross-entropy learning method to optimize PA activation. Numerical results confirmed that the proposed design significantly improved both WPT efficiency and computation performance compared with the conventional antenna scheme. \textcolor{black}{Furthermore, in discrete PA-enabled WPT-MEC systems, TDMA and NOMA exhibit identical performance under coarser PA activation levels, whereas finer activation granularity allows TDMA to achieve superior computation performance. These insights highlight the importance of PA activation flexibility and access scheme selection in the design of WPT-MEC systems.}

\bibliographystyle{IEEEtran}
\bibliography{biblp/bibfilelp}

\end{document}